\documentclass[a4paper,fleqn,usenatbib]{mnras}
\usepackage[T1]{fontenc}
\usepackage{ae,aecompl}
\usepackage{epsfig}
\usepackage{graphicx}
\usepackage{epstopdf}	
\usepackage{amsmath}	
\usepackage{amssymb}
\usepackage{footnote}
\usepackage{longtable}
\usepackage{supertabular,booktabs}	
\usepackage{ulem}
\usepackage{gensymb}

\title{Spectroscopic data of Rb-isoelectronic Zr and Nb ions for astrophysical applications}

\author[Jyoti et al.]{
Jyoti,$^{a}$
{Mandeep Kaur,$^{a}$}
{Bindiya Arora$^{a}$}\thanks{bindiya.phy@gndu.ac.in}
and
{B. K. Sahoo$^{b}$}\thanks{bijaya@prl.res.in}
\\
$^{a}$Department of Physics, Guru Nanak Dev University, Amritsar, Punjab-143005, India\\
$^{b}$Atomic, Molecular and Optical Physics Division, Physical Research Laboratory, Navrangpura, Ahmedabad-380009, India}

\begin{document}

\maketitle

\begin{abstract}
We present high-accuracy spectroscopy data of line strengths, transition probabilities and oscillator strengths for the allowed transitions among
the $nS_{1/2}$, $nP_{1/2,3/2}$ and $n'D_{3/2,5/2}$ states with $n=5$ to $10$ and $n'=4$ to $10$ of the Rb-isoelectronic Zr (Zr IV) and Nb (Nb V) ions. 
They can serve to analyse various astrophysical phenomena undergoing inside the heavenly bodies
containing Zr and Nb elements. Since there is a lack of precise observational and calculated data for the spectroscopic properties in the above ions, 
their accurate determinations are of immense interest. The literature data, that are available only for a few selected low-lying transitions, have 
large discrepancies and they cannot be used reliably for the above purpose. After accounting for electron interactions through random phase 
approximation, Br\"uckner orbitals, structural radiations and normalizations of wave functions in the relativistic many-body methods, we have 
evaluated the electric dipole amplitudes precisely. Combining these values with the observed wavelengths, the above transition properties and 
lifetimes of a number of excited states of the Zr IV and Nb V ions are determined and compared with the literature data.
\end{abstract}

\section{Introduction}

The emission spectra of heavenly bodies has been of great interest since last few decades so as to get the better insight of their atmosphere, 
chemical composition and evolution. A wide variety of analysis have been carried out globally to provide the data for abundance of various spectral
lines in these bodies, however, many of them  are still unknown. A diversity of elements and ions have been observed in these spectra out of which 
zirconium (Zr) and niobium (Nb) are of interest to us, as they play important roles in the decay processes ~\citep{nilsson2010transition,
garcia2007lithium}, which are generally the combination of slow (s) and rapid (r) neutron-capture nucleosynthesis processes and are specifically 
responsible for the detection of spectral lines of lighter elements. {The presence of Nb and its ions has been observed in the atmosphere of three
metal-poor stars HD $209621$, HD $218732$ and HD $232078$, the standard star Arcturus~\citep{niobiumzacs}, Sun and AGB M and MS stars 
~\citep{nilsson2010transition,PALME201415}.} The radionuclide of $^{92}$Nb decays to $^{92}$Zr by neutron-capture process with a half-life of 
$37$ Ma~\citep{holden1989total}. The presence of live $^{92}$Nb in early Solar System was first obtained from the iron meteorite Toluca
~\citep{harper1996evidence}, which further offered a unique opportunity to estimate the timescale of the early Solar System~\citep{IIZUKA2016172}. 
Heavy element stars such as ROA $371$, ROA $5293$, and ROA $3812$ in the globular cluster Omega Centauri consist of a considerable abundance of 
s-process elements; i.e. Rb, Y and Zr ~\citep{vanture1994abundances}. {Besides, the presence of Zr IV lines has been detected in the University 
College London Echelle Spectrograph (UCLES) of a He-rich hot sdB star, LS IV-$14\degree 116$ on the Anglo-Australian Telescope (AAT) 
~\citep{jeffery2011extremely}, in the Far Ultraviolet Spectroscopic Explorer (FUSE) spectra of hot subdwarf B stars (sdB) ~\citep{chayer2006fuse} and
in the UV spectra of hot white dwarfs, G$191$--B$2$B and RE$0503$--$289$ ~\citep{rauch2017viii}.} The spectral properties obtained from measurements 
and calculations are the prerequisites of stellar-atmosphere modeling~\citep{rauch2016stellar}.

The stellar-modeling aspires the determination of chemical abundances and energy transport through a star, which requires the knowledge of reliable 
values of oscillator strengths as well as transition probabilities ~\citep{martin1992fine} of the emission spectra from the elements present in the 
stars.  {Accurate values of absorption oscillator strengths as well as transition probabilities are needed for correctly modeling and analysing 
the stellar intensities of the lines so as to infer fundamental stellar parameters ~\citep{ruffoni2014fe} such as mass \textit{M}, radius \textit{R} 
and luminosity \textit{L} of any star ~\citep{wittkowski2005fundamental}.} The information regarding the oscillator strengths and transition 
probabilities are also useful in the analysis of interstellar and quasar absorption of lines as well as the photospheric abundance of the  considered
element in a star~\citep{rauch2017viii}, the construction of kinetic models of plasma processes and for the investigation of processes in thermonuclear reactor plasma~\citep{glushkov1996calculation,tayal2012breit}. They are also needed for the estimation of the electron collisional rate 
coefficients and photoionization cross-sections so as to explain various scattering phenomena~\citep{griem1974spectral,zeippen1995radiative,
orban2006determination}. The precise evaluation of line strengths is also useful for the assessment of the Stark Broadenings of spectral lines, which 
is pivotal for the analysis of astrophysical phenomena~\citep{alonso2014stark}. Due to this fact, the determination of radiative properties of lines 
present in the emission spectra of stars has become of sheer importance in the astrophysical studies.

To our knowledge, precise observational data of the spectroscopic properties of Rb-isoelectronic Zr (Zr IV) and Nb (Nb V) ions are not available 
in the literature.  Limited theoretical results have been reported on the properties of these ions, but they are only for a few low-lying 
transitions. The data for transition probabilities, oscillator strengths, lifetimes, and branching ratios for Zr IV and Nb V were first provided by 
Lindgard and Nielsen in 1977 by using numerical Coulomb approximation~\citep{lindgaard1977transition}. Later on, in 1979, Migdalek and Baylis studied 
oscillator strengths for the lowest $5S_{1/2}\rightarrow 5P_{1/2,3/2}$ transitions for Rb-isoelectronic series, which included the data for Zr IV and 
Nb V, using relativistic single-configuration Hartree-Fock method and investigated the roles of core polarisation effects for their accurate 
determination~\citep{migdalek1979relativistic}. Karwowski and Szulkin, in 1981, evaluated the ground state energies and also determined the oscillator 
strengths for the $5S \rightarrow 5P$ transitions of Zr IV using the modified relativistic Hartree-Fock method~\citep{karwowski1981relativistic}. Sen 
and Puri had calculated dipole oscillator strengths for these transitions in 1989 using the Local Spin Density approximation and compared them 
with the then available experimental and theoretical data~\citep{sen1989quasi}. A few years later, in 1996, Glushkov et~al. determined oscillator 
strengths for the $5S \rightarrow 5P$, $5P \rightarrow nD$ (n=$4,5$) and $4P \rightarrow 4F$ transitions of Nb V ion using a non-relativistic method 
with the Coulomb screened potential approximation~\citep{glushkov1996calculation}. Zilitis had also evaluated transition probabilities of the 
$5S_{1/2}$--$5P_{1/2,3/2}$, $4D_{3/2}$--$5P_{1/2,3/2}$, and $4D_{3/2}$--$4F_{5/2}$ transitions as well as the lifetimes of the $5P_{1/2}$ 
and $5P_{3/2}$ states of both the Zr IV and Nb V ions using the Dirac-Fock (DF) method in 2007 ~\citep{zilitis2007oscillator} and later, in 2009, he 
extended the calculations for the oscillator strengths of the $4D_{3/2}$--$nP_{1/2,3/2}$ and $4D_{3/2}$--$nF_{5/2}$ transitions of both the ions
\citep{zilitis2009theoretical}. In 2016, transition energies for the $5S$--$5P$, $5P$--$4D$, $4D$--$4F$ and $5D$--$4F$ transitions of these ions were
evaluated by Migdalek on the basis of model potential in the DF method~\citep{migdalek2016relativistic}. Das ~et.al. studied spectroscopic 
properties  of a few Rb-like ions including Zr IV and Nb V ions using the relativistic coupled-cluster method in 2017~\citep{das2017electron}. They 
calculated the matrix elements due to electric and magnetic multipole operators, lifetimes and oscillator strengths for different transitions and 
studied correlation behaviors in the Rb-isoelectronic series from Y III through Tc VII ions.

On account of evidence of presence of Zr and Nb in astrophysical bodies, it is necessary to seek through the radiative data of as many as spectral 
lines. As mentioned above, these data are mainly available on the $ns^{2}S_{1/2} \rightarrow np^{2}P_{1/2,3/2}$, $n=4,5$, $5p^{2}P \rightarrow 
nd^{2}D$, $(n=4,5)$, $4p^{2}P \rightarrow 4f^{2}F$, $4D_{3/2}$--$nP_{1/2,3/2}$ and $4D_{3/2}$--$nF_{5/2}$ transitions only in the above ions. In this 
work, we extend the calculations on the spectroscopic data for a large number of transitions including high-lying lines. In particular, we have 
determined the allowed electric dipole (E1) matrix elements for the transitions among the $nS_{1/2}, nP_{1/2,3/2}$, and $n'D_{3/2,5/2}$ states 
with $n=5$ to $10$ and $n'=4$ to $10$ of the Zr IV and Nb V ions except for a few transitions which were not gauge invariant. Combining these data 
with the observed transition wavelengths, we have estimated the oscillator strengths, transition probabilities and radiative lifetimes of a number 
of excited states precisely. Moreover, we have calculated energies of a few low-lying and excited states for both the ions and compared them 
with the values listed in the National Institute of Science and Technology atomic database (NIST AD) ~\citep{ralchenko2008nist}. The aforementioned 
data can be immensely useful for analysing astrophysical processes; especially in the metal-poor stars HD $209621$, HD $218732$ and HD $232078$, 
Arcturus and white dwarfs.

The paper is organized as follows: In Sec. \ref{2}, we provide the theoretical formulae for the E1 matrix elements, transition probabilities, 
oscillator strengths and lifetime of the atomic states. Sec. \ref{4} discusses all the acquired data from present work and compares with the 
previously reported values, while findings are concluded in Sec. \ref{5}. All the results are given in atomic units (a.u.) unless stated otherwise.

\section{Theoretical aspects}
\label{2}

\subsection{Formulae for spectroscopic quantities}

 Transitions among the atomic states are dominantly driven by the E1 channel when allowed. Here, we analyze the transition properties only due to
the E1 channel as electrons can decay from the excited states of the ions of our interest mainly through this channel. The transition probability 
due to the E1 channel ($A^{E1}_{vk}$) from an upper level $|\Psi(v)\rangle$ to a lower level $|\Psi(k)\rangle$ with the 
corresponding angular momentum $J_v$ and $J_k$ respectively is given (in $s^{-1}$) by \citep{kelleher2008atomic}
\begin{equation}
A^{E1}_{vk}= \frac{2}{3}\alpha c \pi \sigma \left(\frac{\alpha\sigma}{R_{\infty}}\right)^{2} \frac{S^{E1}}{g_{v}},
\end{equation}
{where $R_{\infty}=\frac{\alpha^{2} m_{e} c}{2h}$ is the Rydberg constant with 
the Planck's constant $h$ and mass of electron $m_e$, $\alpha$ is fine structure constant $\alpha  = \frac{e^{2}}{4 \pi \epsilon_{0} \hbar c}$, $c$ is the speed of light}, 
$\sigma=E_{v}-E_{k}$ is the excitation energy of the transition, $S^{E1}$ is the line strength and $g_v= 2J_v +1$. Here, $S^{E1}=|\langle J_{v}||
\textbf{D}||J_k \rangle|^{2}$ \citep{nahar1995atomic} with the E1 operator $\textbf{D}=\Sigma_{j} \textbf{d}_{j}=-e \Sigma_{j} \textbf{r}_{j}$ for 
the $j^{th}$ electron being at position $\textbf{r}_{j}$. On substituting the values of fundamental constants {as $\alpha  = 7.297352\times10^{-3}$, $R_{\infty} =1.0973731\times 10^5$ cm$^{-1}$ and
$c=29979245800$ cm s$^{-1}$~\citep{mohr},} the above formula yields 
\citep{kelleher2008atomic}
\begin{equation}
A^{E1}_{vk}=\frac{2.0261269\times10^{18} }{\lambda^{3}g_{v}} S^{E1},
\end{equation}
where $\lambda$ is the wavelength of transition in $\text{\normalfont\AA}$. 

The absorption oscillator strengths of the allowed transitions ($f^{E1}_{kv}$) are very useful in the astrophysical analyses, which can be 
determined from the {lower $|\Psi(k)\rangle$} level to the {upper $|\Psi(v)\rangle$ level} using its transition probability which follows as \citep{kelleher2008atomic,sobelman1979angular,kaur2020radiative}
\begin{eqnarray}
f_{kv}^{E1}& =& \left(\frac{R_{\infty}}{2c\alpha^{3}\pi} \right)\frac{g_{v}}{g_{k}} \times \frac{A_{vk}^{E1}}{\sigma^{2}} \nonumber \\
  &=& \frac{1}{3\alpha}\left(\frac{\alpha\sigma}{R_{\infty}} \right) \times \frac{S^{E1}}{g_{k}} \nonumber \\
  &=&\frac{303.756}{g_{k}\lambda}\times S^{E1}.
\end{eqnarray}

The radiative lifetime ($\tau$) of the excited state{$|\Psi(v)\rangle$} can also be obtained by taking reciprocal of the total transition 
probabilities {of all the lower possible transitions from that level} \citep{qin2019energy,kaur2020radiative}; i.e.
\begin{equation}
\tau_{v}=\frac{1}{\sum_k A_{vk}} ,
\end{equation}
where sum over $k$ denotes all possible states ($|\Psi(k)\rangle$) to which allowed transitions from $|\Psi(v)\rangle$ are possible.

\subsection{Methods for calculations}

The presence of the two-body electromagnetic interactions among the electrons within an atomic system poses challenge to solve atomic wave functions
accurately. In a typical approach, atomic wave functions are obtained by calculating them using a mean-field approach then incorporating the electron
correlation effects due to the neglected interactions systematically. These correlation effects are classified into various physical effects such 
as core-polarization and pair-correlation effects, which behave differently for accurate description of wave functions depending upon the electronic 
{configurations} of atomic systems. Large differences among the previous {calculations} of transition properties in the considered ions using a variety of 
many-body methods \citep{lindgaard1977transition,migdalek1979relativistic,karwowski1981relativistic,glushkov1996calculation,zilitis2007oscillator,
das2017electron} suggest that the roles of the electron correlation effects are significant and they should be accounted for meticulously in order to
obtain reliable results. In view of this, we include the electron correlations due to the core-polarization effects through random-phase 
approximation (RPA), pair-correlation effects through the Br\"uckner orbitals (BOs) and their couplings through the structural radiations (SRs) in 
the determination of atomic wave functions. Corrections in the results due to normalization of the wave functions (Norms) are also estimated 
explicitly. A brief description of the procedures adopted to incorporate the above physical effects in the calculations is given below. 

We first evaluate the mean-field wave function ($|\Phi_0 \rangle$) of the $[4p^6]$ configuration of the considered ions using the DF method 
in which the atomic Hamiltonian is expressed as $H=H_0+V_I$ with the DF Hamiltonian $H_0$ and residual interaction $V_I$, given in atomic 
units (a.u.) by 
\begin{equation}
H_0=\sum_i \epsilon_ia_i^{\dag}a_i \label{nopair}
\end{equation}
and
\begin{equation}
V_I=\frac{1}{2}\sum_{ijkl} g_{ijkl}a_i^{\dag}a_j^{\dag}a_la_k - \sum_{ij} u_{ij}^{DF} a_i^{\dag} a_j,
\end{equation} 
where the sums over $i,j,k$ and $l$ include all electron orbitals, $\epsilon_i$ are the eigenvalues of the one-electron DF orbitals, $g_{ijkl}$ is 
the two-body matrix element of the Coulomb interaction ($\frac{1}{r_{ij}}$) and $u_{ij}^{DF}$ is the DF potential \citep{Blundell1987,Johnson1996,
safronova2005excitation}. The working DF wave functions ($|\Phi_v \rangle$) of the interested states of both Zr IV and Nb V are obtained by appending the 
respective valence orbital $v$ to $|\Phi_0 \rangle$ for the configurations $[4p^6]v$; i.e. $|\Phi_v \rangle=a_v^{\dag} |\Phi_0 \rangle$. The choice
this $V^{N-1}$ DF potential with ($N$ number of electrons of the system) is to facilitate for calculating as many as states having common closed core 
$[4p^6]$. The neglected core-valence effects at this level are included as corrections in the next level.

The corrections over the DF wave functions due to the electron correlation effects are estimated using the perturbative analysis of $V_I$ by 
expressing the exact wave function of the state ($|\Psi_v \rangle$) in the relativistic many-body perturbation theory (RMBPT) analysis as
\begin{eqnarray}
 |\Psi_v \rangle = |\Phi_v \rangle + |\Phi_v^{(1)} \rangle + |\Phi_v^{(2)} \rangle + \cdots 
\end{eqnarray}
and its energy $(E_v$) as
\begin{eqnarray}
 E_v = E_v^{(0)} + E_v^{(1)}+ E_v^{(2)} + \cdots,
\end{eqnarray}
where superscripts $k=1,2$ etc. denote order of {perturbation} and the zeroth-order energy is $E_v^{(0)}= \sum_k^N \epsilon_k$. After obtaining the
wave functions, the E1 matrix element between the states $|\Psi_v \rangle$ and $|\Psi_w \rangle$ is calculated as 
\begin{equation}
D_{wv}=\frac{\left\langle\Psi_w|D|\Psi_v\right\rangle}{\sqrt{\left\langle\Psi_w|\Psi_w\right\rangle \left\langle\Psi_v|\Psi_v\right\rangle}} .  \label{matel}
\end{equation}
As mentioned above contributions from the perturbative corrections are categorized into RPA, BO, SR and Norm contributions by
expressing \citep{Blundell1987,Johnson1996}
\begin{eqnarray}
 D_{wv}= D^{DF}_{wv} + D^{RPA}_{wv}+D^{BO}_{wv}+D^{SR}_{wv}+D^{Norm}_{wv} ,
\end{eqnarray}
where $D^{DF}_{wv} \equiv d_{wv}= \langle \phi_w | d | \phi_v \rangle$ with the DF single particle wave functions $|\phi_v \rangle$ and $|\phi_w\rangle$.
Since core-polarization effects contribute significantly, they are included through RPA self-consistently to all-orders by defining {\citep{johnson2007atomic,Johnson1996}}:
\begin{eqnarray}
D_{wv}^{RPA} = \sum_{\zeta=1}^{\infty} \sum_{na} \left [ \frac{d_{an}^{(\zeta)} \tilde{g}_{wnva}}{\epsilon_v+\epsilon_a-\epsilon_n-\epsilon_w} 
 + \frac{ \tilde{g}_{wavn} d^{(\zeta)}_{na}}{\epsilon_w+\epsilon_a-\epsilon_n-\epsilon_v} \right ] \nonumber\\ \label{2nd}
\end{eqnarray}
for $\tilde{g}_{ijkl} =g_{ijkl}- g_{ijlk}$ and superscript $\zeta$ represents iteration number with
\begin{eqnarray}
d_{an}^{(\zeta)}&=&d_{an}+\sum_{bm} \left [ \frac{d_{bm}^{(\zeta-1)} \tilde{g}_{amnb} } {\epsilon_v + \epsilon_{b}-\epsilon_{m}-\epsilon_w}
+ \frac{\tilde{g}_{abnm} d_{mb}^{(\zeta-1)} }{\epsilon_w+\epsilon_{b}-\epsilon_{m}-\epsilon_v} \right ], \nonumber 
\end{eqnarray}
and
\begin{eqnarray}
d_{na}^{(\zeta)}&=&d_{na}+\sum_{bm} \left [ \frac{d_{bm}^{(\zeta-1)} \tilde{g}_{nmab}}{\epsilon_v+\epsilon_{b}-\epsilon_{m}-\epsilon_w}
+ \frac{\tilde{g}_{nbam} d_{mb}^{(\zeta-1)} }{\epsilon_w+\epsilon_{b}-\epsilon_{m}-\epsilon_v} \right ] . \nonumber
\end{eqnarray}
The initial values correspond to $d^{(0)}_{an}=0$ and $d^{(1)}_{an}=d_{an}$. The $a,b,...$, $m,n,...$ and $w,v$ indices in the subscripts denote for 
the occupied, virtual and valence orbitals, respectively.

{The leading-order electron correlation contributions through BO and SR arise at the third-order perturbation level. We include these contributions 
using the procedure followed by Johnson et~al. in \citep{Johnson1996} in the present calculations. The Norm contributions have been estimated by 
adopting the approach discussed in ~\citep{Blundell1987} and approximating the atomic wave function at the second-order perturbation theory.}

It is obvious from the above discussion that our procedure incorporates various physical effects due to the electron correlation effects that are 
complete through the third-order perturbation and core-polarization effects to all-orders. To verify reliability in the calculations of the E1 
matrix elements and estimate their uncertainties, we use both the length (L) and velocity gauge (V) expressions in the relativistic form (e.g. see
~\citep{kaur2020radiative}).  The differences in the results from both the gauge forms can be safely used as the maximum 
uncertainties associated with our calculated E1 matrix elements. Since calculations with length gauge expression converge faster with respect to 
the number of configurations, we believe that results from the length gauge expression are more reliable. Thus, we consider these results as the 
central values. In a few transitions, it is noticed that the length gauge and velocity gauge results differ by more than 50\%. In such cases,  
it is possible to improve calculations by considering higher-order contributions through BO and SR effects, but they are computationally more 
expensive. Therefore, we do not endorse the results obtained using above method for such few cases and do not present them in this work.
\begin{table}
\caption{\label{tab10} Our calculated energy values ($cm^{-1}$) for few low lying and excited states using RMBPT method for Zr IV and Nb V along with their comparison with experimental energies provided in NIST AD ~\citep{ralchenko2008nist}. Ground state is taken at $0$ cm$^{-1}$.}
\begin{center}
\begin{tabular}{@{\extracolsep\fill}ccccc@{}}
\hline
State & \multicolumn{2}{c}{Zr IV} & \multicolumn{2}{c}{Nb V}\\
 & Energy ($cm^{-1}$) & NIST ($cm^{-1}$) & Energy ($cm^{-1}$) & NIST ($cm^{-1}$)\\
\hline
$4D_{5/2}$ & $1109.05$ & $1250.70$ & $1741.83$ & $1867.40$\\
$5S_{1/2}$ & $34572.42$ & $38258.35$ & $72113.24$ & $75929.60$\\
$5P_{1/2}$ & $76313.25$ & $81976.50$ & $123581.01$ & $129195.20$\\
$5P_{3/2}$ & $78607.03$ & $84461.35$ & $126972.64$ & $132800.00$\\
$5D_{3/2}$ & $138820.28$ & $146652.40$ & $203904.11$ & $211694.00$\\
$5D_{5/2}$ & $139158.53$ & $147002.46$ & $204434.63$ & $212238.40$\\
$6S_{1/2}$ & $143994.52$ & $152513.00$ & $219451.86$ & $228496.30$\\
$6P_{1/2}$ & $160898.86$ & $169809.71$ & $241153.11$ & $250506.50$\\
$6P_{3/2}$ & $161848.63$ & $170815.11$ & $242602.26$ & $252023.3$\\
$6D_{3/2}$ & $188289.90$ & $197765.10$ & $277261.63$ & $287163.60$\\
$6D_{5/2}$ & $188450.91$ & $197930.43$ & $277515.77$ & $287425.60$\\
$7S_{1/2}$ & $190309.74$ & $200123.69$ & $284180.44$ & $294736.00$\\
$7P_{1/2}$ & $198869.74$ & $208783.36$ & $295423.65$ & $305986.50$\\
$7P_{3/2}$ & $199358.47$ & $ 209297.66$ & $296185.80$ & $306788.10$\\
$7D_{3/2}$ & $213460.26$ &  & $314942.71$ & $325705.70$\\
$7D_{5/2}$ & $213550.09$ &  & $315085.29$ & $325854.80$\\
$8S_{1/2}$ & $214494.56$ & $224813.48$ & $318702.16$ & $329884.20$\\
\hline
\end{tabular}
\end{center}
\end{table}

\newpage
\clearpage
\onecolumn
\begin{longtable}{@{\extracolsep\fill}rllllllll@{}}
\caption{\label{tab3}The line strengths ($S_{vk}$) (in a.u.) from both the L and V gauge expressions, wavelengths ($\lambda$) (in $\text{\normalfont\AA}$), transition probabilities ($A_{vk}$) in ($s^{-1}$) and absorption oscillator strengths ($f_{kv}$) for the Zr IV ion through E1 decay channel are presented in this table. Values in square brackets represent the order of 10. Uncertainties are given in parentheses.}\\
\hline
Upper State(v) & Lower State(k) & $\lambda$ (in $\text{\normalfont\AA}$) & \multicolumn{2}{c}{$S_{vk}$ (in a.u.)} & $A_{L_{vk}}$(in $s^{-1}$) & $f_{L_{kv}}$\\ 
 & & & L & V & &\\
 \hline
$5P_{1/2}$	&	$4D_{3/2}$	&	$1219.86$	&	$2.04[0]$	&	$2.37[0]$	&	$1.14(18)[9]$	&	$1.27(20)[-1]$\\
$5P_{1/2}$	&	$5S_{1/2}$	&	$2287.38$	&	$4.69[0]$	&	$5.06[0]$	&	$3.97(30)[8]$	&	$3.12(24)[-1]$\\
												
$5P_{3/2}$	&	$4D_{3/2}$	&	$1183.97$	&	$3.93[-1]$	&	$4.55[-1]$	&	$1.20(18)[8]$	&	$2.52(38)[-2]$\\
$5P_{3/2}$	&	$4D_{5/2}$	&	$1201.77$	&	$3.64[0]$	&	$4.21[0]$	&	$1.06(16)[9]$	&	$1.54(23)[-1]$\\
$5P_{3/2}$	&	$5S_{1/2}$	&	$2164.36$	&	$9.40[0]$	&	$1.02[1]$	&	$4.70(39)[8]$	&	$6.60(54)[-1]$\\
												
$5D_{3/2}$	&	$5P_{1/2}$	&	$1546.17$	&	$1.14[1]$	&	$1.20[1]$	&	$1.56(8)[9]$ &	$1.12(6)[0]$\\
$5D_{3/2}$	&	$5P_{3/2}$	&	$1607.95$	&	$2.40[0]$	&	$2.51[0]$	&	$2.92(14)[8]$	&	$1.13(5)[-1]$\\
												
$5D_{5/2}$	&	$5P_{3/2}$	&	$1598.95$	&	$2.13[1]$	&	$2.25[1]$	&	$1.76(9)[9]$	&	$1.01(5)[0]$\\
												
$6S_{1/2}$	&	$5P_{1/2}$	&	$1417.71$	&	$1.82[0]$	&	$1.90[0]$	&	$6.47(27)[8]$	&	$1.95(8)[-1]$\\
$6S_{1/2}$	&	$5P_{3/2}$	&	$1469.47$	&	$4.06[0]$	&	$4.21[0]$	&	$1.30(5)[9]$	&	$2.10(8)[-1]$\\
												
{$6P_{1/2}$}	&	$4D_{3/2}$	&	$588.89$	&	$6.51[-2]$	&	$7.46[-2]$	&	$3.23(46)[8]$	&	 {$8.4(1.2)[-3]$}\\
$6P_{1/2}$	&	$5S_{1/2}$	&	$760.16$	&	$4.33[-2]$	&	$4.84[-2]$	&	$1.00(11)[8]$	&	$8.66(10)[-3]$\\
$6P_{1/2}$	&	$5D_{3/2}$	&	$4318.29$	&	$1.57[1]$	&	$1.63[1]$	&	$1.97(8)[8]$	&	$2.75(11)[-1]$\\
$6P_{1/2}$	&	$6S_{1/2}$	&	$5781.45$	&	$1.83[1]$	&	$1.91[1]$	&	$9.61(40)[7]$	&	$4.82(20)[-1]$\\
												
$6P_{3/2}$	&	$4D_{3/2}$	&	$585.42$	&	$1.42[-2]$	&	$1.61[-2]$	&	$3.59(46)[7]$	&	$1.84(24)[-3]$\\
$6P_{3/2}$	&	$4D_{5/2}$	&	$589.75$	&	$1.29[-1]$	&	$1.47[-1]$	&	$3.20(42)[8]$	&	$1.11(15)[-2]$\\
$6P_{3/2}$	&	$5S_{1/2}$	&	$754.39$	&	$5.35[-2]$	&	$6.10[-2]$	&	$6.31(86)[7]$	&	$1.08(15)[-2]$\\
$6P_{3/2}$	&	$5D_{3/2}$	&	$4138.61$	&	$3.00[0]$	&	$3.14[0]$	&	$2.15(10)[7]$	&	$5.51(25)[-2]$\\
$6P_{3/2}$	&	$5D_{5/2}$	&	$4199.45$	&	$2.75[1]$	&	$2.88[1]$	&	$1.88(8)[8]$	&	$3.32(15)[-1]$\\
$6P_{3/2}$	&	$6S_{1/2}$	&	$5463.85$	&	$3.64[1]$	&	$3.80[1]$	&	$1.13(5)[8]$	&	$1.01(5)[0]$\\
												
$6D_{3/2}$	&	$5P_{1/2}$	&	$863.64$	&	$2.84[-1]$	&	$2.86[-1]$	&	$2.24(2)[8]$	&	$5.00(4)[-2]$\\
$6D_{3/2}$	&	$5P_{3/2}$	&	$882.58$	&	$5.01[-2]$	&	$5.03[-2]$	&	$3.69(1)[7]$	&	$4.31(1)[-3]$\\
$6D_{3/2}$	&	$6P_{1/2}$	&	$3577.13$	&	$3.40[1]$	&	$3.50[1]$	&	$3.76(11)[8]$	&	$1.44(4)[0]$\\
$6D_{3/2}$	&	$6P_{3/2}$	&	$3710.58$	&	$7.17[0]$	&	$7.36[0]$	&	$7.11(19)[7]$	&	$1.47(4)[-1]$\\
												
$6D_{5/2}$	&	$5P_{3/2}$	&	$881.30$	&	$4.73[-1]$	&	$4.76]-1]$	&	$2.34(1)[8]$	&	$4.08(2)[-2]$\\
$6D_{5/2}$	&	$6P_{3/2}$	&	$3687.95$	&	$6.39[1]$	&	$6.56[1]$	&	$4.30(12)[8]$	&	$1.32(4)[0]$\\
												
$7S_{1/2}$	&	$5P_{1/2}$	&	$846.40$	&	$1.73[-1]$	&	$1.74[-1]$	&	$2.89(2)[8]$	&	$3.10(3)[-2]$\\
$7S_{1/2}$	&	$5P_{3/2}$	&	$864.59$	&	$3.62[-1]$	&	$3.63[-1]$	&	$5.68(2)[8]$	&	$3.18(1)[-2]$\\
$7S_{1/2}$	&	$6P_{1/2}$	&	$3298.81$	&	$6.32[0]$	&	$6.59[0]$	&	$1.78(7)[8]$	&	$2.91(12)[-1]$\\
$7S_{1/2}$	&	$6P_{3/2}$	&	$3411.97$	&	$1.40[1]$	&	$1.45[1]$	&	$3.57(13)[8]$	&	$3.12(12)[-1]$\\
												
$7P_{1/2}$	&	$4D_{3/2}$	&	$478.97$	&	$1.47[-2]$	&	$1.73[-2]$	&	$1.35(23)[8]$	&	$2.33(39)[-3]$\\
$7P_{1/2}$	&	$5S_{1/2}$	&	$586.42$	&	$2.22[-2]$	&	$2.38[-2]$	&	$1.11(8)[8]$	&	$5.74(41)[-3]$\\
$7P_{1/2}$	&	$5D_{3/2}$	&	$1609.50$	&	$2.29[-1]$	&	$2.18[-1]$	&	$5.57(28)[7]$	&	$1.08(5)[-2]$\\
$7P_{1/2}$	&	$6S_{1/2}$	&	$1777.13$	&	$3.29[-2]$	&	$3.84[-2]$	&	$5.93(96)[6]$	&	$2.81(46)[-3]$\\
$7P_{1/2}$	&	$6D_{3/2}$	&	$9075.84$	&	$5.01[1]$	&	$5.20[1]$	&	$6.79(26)[7]$	&	$4.19(16)[-1]$\\
$7P_{1/2}$	&	$7S_{1/2}$	&	$11547.78$	&	$4.69[1]$	&	$4.83[1]$	&	$3.09(9)[7]$	&	$6.17(17)[-1]$\\
												
$7P_{3/2}$	&	$4D_{3/2}$	&	$477.79$	&	$3.36[-3]$	&	$3.87[-3]$	&	$1.56(23)[7]$	&	$5.35(77)[-4]$\\
$7P_{3/2}$	&	$4D_{5/2}$	&	$480.66$	&	$3.08[-2]$	&	$3.56[-2]$	&	$1.40(21)[8]$	&	$3.24(49)[-3]$\\
$7P_{3/2}$	&	$5S_{1/2}$	&	$584.66$	&	$3.27[-2]$	&	$3.55[-2]$	&	$8.28(70)[7]$	&	$8.48(71)[-3]$\\
$7P_{3/2}$	&	$5D_{3/2}$	&	$1596.29$	&	$5.21[-2]$	&	$5.01[-2]$	&	$6.49(26)[6]$	&	$2.48(10)[-3]$\\
$7P_{3/2}$	&	$5D_{5/2}$	&	$1605.26$	&	$4.58[-1]$	&	$4.39[-1]$	&	$5.61(23)[7]$	&	$1.44(6)[-2]$\\
$7P_{3/2}$	&	$6S_{1/2}$	&	$1761.04$	&	$2.24[-2]$	&	$2.88[-2]$	&	$2.08(56)[6]$	&	$1.94(52)[-3]$\\
$7P_{3/2}$	&	$6D_{3/2}$	&	$8671.10$	&	$9.62[0]$	&	$1.00[1]$	&	$7.47(33)[6]$	&	$8.42(37)[-2]$\\
$7P_{3/2}$	&	$6D_{5/2}$	&	$8797.22$	&	$8.79[1]$	&	$9.18[1]$	&	$6.54(28)[7]$	&	$5.06(22)[-1]$\\
$7P_{3/2}$	&	$7S_{1/2}$	&	$10900.41$	&	$9.27[1]$	&	$9.56[1]$	&	$3.62(12)[7]$	&	$1.29(4)[0]$\\
												
$7D_{3/2}$	&	$5P_{1/2}$	&	$729.14$	&	$4.52[-2]$	&	$4.39[-2]$	&	$5.91(17)[7]$	&	$9.42(28)[-3]$\\
$7D_{3/2}$	&	$5P_{3/2}$	&	$741.55$	&	$7.23[-3]$	&	$6.96[-3]$	&	$8.97(34)[6]$	&	$7.40(28)[-4]$\\
$7D_{3/2}$	&	$6P_{1/2}$	&	$1902.54$	&	$1.21[0]$	&	$1.21[0]$	&	$8.88(3)[7]$	&	$9.64(3)[-2]$\\
 {$7D_{3/2}$}	&	$6P_{3/2}$	&	$1937.55$	&	$2.22[-1]$	&	$2.22[-1]$	&	$1.5449(7)[7]$ &	$8.695(4)[-3]$\\
$7D_{3/2}$	&	$7P_{1/2}$	&	$6853.77$	&	$7.55[1]$	&	$7.71[1]$	&	$1.19(2)[8]$	&	$1.67(3)[0]$\\
$7D_{3/2}$	&	$7P_{3/2}$	&	$7091.30$	&	$1.60[1]$	&	$1.63[1]$	&	$2.27(4)[7]$	&	$1.71(3)[-1]$\\
												
$7D_{5/2}$	&	$5P_{3/2}$	&	$741.05$	&	$7.07[-2]$	&	$6.84[-2]$	&	$5.87(19)[7]$	&	$7.25(24)[-3]$\\
$7D_{5/2}$	&	$6P_{3/2}$	&	$1934.18$	&	$2.06[0]$	&	$2.06[0]$	&	$9.59(1)[7]$	&	$8.07(1)[-2]$\\
$7D_{5/2}$	&	$7P_{3/2}$	&	$7046.41$	&	$1.42[2]$	&	$1.45[2]$	&	$1.37(3)[8]$	&	$1.53(3)[0]$\\
												
$8S_{1/2}$	&	$5P_{1/2}$	&	$700.10$	&	$5.41[-2]$	&	$5.37[-2]$	&	$1.60(1)[8]$	&	$1.17(1)[-2]$\\
$8S_{1/2}$	&	$5P_{3/2}$	&	$712.49$	&	$1.12[-1]$	&	$1.10[-1]$	&	$3.13(4)[8]$	&	$1.19(1)[-2]$\\
$8S_{1/2}$	&	$6P_{1/2}$	&	$1818.06$	&	$5.04[-1]$	&	$5.15[-1]$	&	$8.50(18)[7]$	&	$4.21(9)[-2]$\\
$8S_{1/2}$	&	$6P_{3/2}$	&	$1851.91$	&	$1.05[0]$	&	$1.06[0]$	&	$1.67(3)[8]$	&	$4.30(7)[-2]$\\
$8S_{1/2}$	&	$7P_{1/2}$	&	$6238.26$	&	$1.60[1]$	&	$1.66[1]$	&	$6.66(27)[7]$	&	$3.88(16)[-1]$\\
$8S_{1/2}$	&	$7P_{3/2}$	&	$6445.03$	&	$3.52[1]$	&	$3.64[1]$	&	$1.33(5)[8]$	&	$4.14(15)[-1]$\\
												
 {$8P_{1/2}$}	&	$4D_{3/2}$	&	$455.75$	&	$5.34[-3]$	&	$6.48[-3]$	&	 {$5.7(1.2)[7]$}	&	 {$8.9(1.8)[-4]$}\\
$8P_{1/2}$	&	$5S_{1/2}$	&	$540.99$	&	$1.24[-2]$	&	$1.30[-2]$	&	$7.91(41)[7]$	&	$3.47(18)[-3]$\\
$8P_{1/2}$	&	$5D_{3/2}$	&	$1240.73$	&	$5.17[-2]$	&	$4.77[-2]$	&	$2.74(21)[7]$	&	$3.16(25)[-3]$\\
 {$8P_{1/2}$}	&	$6S_{1/2}$	&	$1325.84$	&	$2.14[-2]$	&	$2.38[-2]$	&	 {$9.3(1.0)[6]$}	&	$2.45(27)[-3]$\\
$8P_{1/2}$	&	$6D_{3/2}$	&	$3212.52$	&	$5.38[-1]$	&	$5.14[-1]$	&	$1.64(7)[7]$	&	$1.27(6)[-2]$\\
 {$8P_{1/2}$}	&	$7S_{1/2}$	&	$3435.43$	&	$2.81[-2]$	&	$3.37[-2]$	&	 {$7.0(1.3)[5]$}	&	$1.24(24)[-3]$\\
$8P_{1/2}$	&	$7D_{3/2}$	&	$16784.45$	&	$1.19[2]$	&	$1.24[2]$	&	$2.55(11)[7]$	&	$5.38(23)[-1]$\\
 {$8P_{1/2}$}	&	$8S_{1/2}$	&	$20310.35$	&	$9.89[1]$	&	$1.01[2]$	&	$1.20(2)[7]$	&	$7.40(15)[-1]$\\
												
 {$8P_{3/2}$}	&	$4D_{3/2}$	&	$455.16$	&	$1.28[-3]$	&	$1.51[-3]$	&	 {$6.9(1.2)[6]$}	&	$2.14(37)[-4]$\\
$8P_{3/2}$	&	$4D_{5/2}$	&	$457.47$	&	$1.17[-2]$	&	$1.39[-2]$	&	 {$6.2(1.1)[7]$}	&	$1.30(23)[-3]$\\
$8P_{3/2}$	&	$5S_{1/2}$	&	$540.15$	&	$1.92[-2]$	&	$2.04[-2]$	&	$6.16(39)[7]$	&	$5.39(3)[-3]$\\
$8P_{3/2}$	&	$5D_{3/2}$	&	$1236.33$	&	$1.20[-2]$	&	$1.12[-2]$	&	$3.21(21)[6]$	&	$7.35(48)[-4]$\\
$8P_{3/2}$	&	$5D_{5/2}$	&	$1241.52$	&	$1.05[-1]$	&	$9.78[-2]$	&	$2.77(19)[7]$	&	$4.27(29)[-3]$\\
$8P_{3/2}$	&	$6S_{1/2}$	&	$1320.82$	&	$2.46[-2]$	&	$2.84[-2]$	&	$5.41(80)[6]$	&	$2.83(42)[-3]$\\
$8P_{3/2}$	&	$6D_{3/2}$	&	$3183.19$	&	$1.26[-1]$	&	$1.22[-1]$	&	$1.98(7)[6]$	&	$3.00(10)[-3]$\\
 {$8P_{3/2}$}	&	$6D_{5/2}$	&	$3199.59$	&	$1.10[0]$	&	$1.06[0]$	&	$1.69(6)[7]$	&	$1.73(6)[-2]$\\
$8P_{3/2}$	&	$7S_{1/2}$	&	$3401.92$	&	$6.29[-3]$	&	$1.02[-2]$	&	 {$8.1(4.4)[4]$}	&	 {$2.8(1.5)[-4]$}\\
$8P_{3/2}$	&	$7D_{3/2}$	&	$16013.68$	&	$2.29[1]$	&	$2.39[1]$	&	$2.82(13)[6]$	&	$1.09(5)[-1]$\\
$8P_{3/2}$	&	$7D_{5/2}$	&	$16247.40$	&	$2.09[2]$	&	$2.18[2]$	&	$2.47(11)[7]$	&	$6.51(29)[-1]$\\
$8P_{3/2}$	&	$8S_{1/2}$	&	$19192.52$	&	$1.95[2]$	&	$1.99[2]$	&	$1.40(3)[7]$	&	$1.54(4)[0]$\\
												
$8D_{3/2}$	&	$5P_{1/2}$	&	$658.67$	&	$1.31[-2]$	&	$1.23[-2]$	&	$2.32(14)[7]$	&	$3.02(18)[-3]$\\
$8D_{3/2}$	&	$5P_{3/2}$	&	$668.77$	&	$1.93[-3]$	&	$1.79[-3]$	&	$3.26(24)[6]$	&	$2.19(16)[-4]$\\
$8D_{3/2}$	&	$6P_{1/2}$	&	$1487.30$	&	$2.45[-1]$	&	$2.41[-1]$	&	$3.78(7)[7]$	&	$2.51(4)[-2]$\\
$8D_{3/2}$	&	$6P_{3/2}$	&	$1508.61$	&	$4.26[-2]$	&	$4.17[-2]$	&	$6.29(14)[6]$	&	$2.15(5)[-3]$\\
$8D_{3/2}$	&	$7P_{1/2}$	&	$3417.04$	&	$2.97[0]$	&	$2.99[0]$	&	$3.78(2)[7]$	&	$1.32(1)[-1]$\\
$8D_{3/2}$	&	$7P_{3/2}$	&	$3475.08$	&	$5.56[-1]$	&	$5.57[-1]$	&	$6.71(2)[6]$	&	 {$1.215(3)[-2]$}\\
$8D_{3/2}$	&	$8P_{1/2}$	&	$11472.29$	&	$1.45[2]$	&	$1.47[2]$	&	$4.85(7)[7]$	&	$1.91(3)[0]$\\
$8D_{3/2}$	&	$8P_{3/2}$	&	$11862.56$	&	$3.07[1]$	&	$3.11[1]$	&	$9.31(12)[6]$	&	$1.96(2)[-1]$\\
												
$8D_{5/2}$	&	$5P_{3/2}$	&	$668.52$	&	$1.95[-2]$	&	$1.83[-2]$	&	$2.21(15)[7]$	&	$2.22(15)[-3]$\\
$8D_{5/2}$	&	$6P_{3/2}$	&	$1507.35$	&	$4.02[-1]$	&	$3.93[-1]$	&	$3.96(8)[7]$	&	$2.02(4)[-2]$\\
$8D_{5/2}$	&	$7P_{3/2}$	&	$3468.39$	&	$5.12[0]$	&	$5.13[0]$	&	$4.14(1)[7]$	&	 {$1.121(3)[-1]$}\\
$8D_{5/2}$	&	$8P_{3/2}$	&	$11785.04$	&	$2.73[2]$	&	$2.76[2]$	&	$5.63(7)[7]$	&	$1.76(2)[0]$\\
												
$9S_{1/2}$	&	$5P_{1/2}$	&	$655.99$	&	$2.49[-2]$	&	$2.45[-2]$	&	$8.94(15)[7]$	&	$5.77(10)[-3]$\\
$9S_{1/2}$	&	$5P_{3/2}$	&	$666.01$	&	$5.10[-2]$	&	$5.00[-2]$	&	$1.75(4)[8]$	&	$5.81(12)[-3]$\\
$9S_{1/2}$	&	$6P_{1/2}$	&	$1473.70$	&	$1.46[-1]$	&	$1.48[-1]$	&	$4.64(5)[7]$	&	$1.51(2)[-2]$\\
$9S_{1/2}$	&	$6P_{3/2}$	&	$1494.62$	&	$3.00[-1]$	&	$3.01[-1]$	&	$9.09(5)[7]$	&	$1.52(1)[-2]$\\
$9S_{1/2}$	&	$7P_{1/2}$	&	$3346.08$	&	$1.16[0]$	&	$1.19[0]$	&	$3.12(8)[7]$	&	$5.24(14)[-2]$\\
$9S_{1/2}$	&	$7P_{3/2}$	&	$3401.71$	&	$2.39[0]$	&	$2.44[0]$	&	$6.15(13)[7]$	&	$5.33(12)[-2]$\\
$9S_{1/2}$	&	$8P_{1/2}$	&	$10709.75$	&	$3.34[1]$	&	$3.46[1]$	&	$2.75(11)[7]$	&	$4.73(18)[-1]$\\
$9S_{1/2}$	&	$8P_{3/2}$	&	$10813.09$	&	$7.34[1]$	&	$7.58[1]$	&	$5.88(19)[7]$	&	$5.15(17)[-1]$\\
												
 {$9P_{1/2}$}	&	$4D_{3/2}$	&	$431.37$	&	$2.45[-3]$	&	$3.05[-3]$	&	$3.09(71)[7]$	&	$4.31(99)[-4]$\\
$9P_{1/2}$	&	$5S_{1/2}$	&	$506.98$	&	$7.59[-3]$	&	$7.90[-3]$	&	$5.90(24)[7]$	&	$2.27(9)[-3]$\\
$9P_{1/2}$	&	$5D_{3/2}$	&	$1075.28$	&	$2.05[-2]$	&	$1.87[-2]$	&	$1.67(15)[7]$	&	$1.45(13)[-3]$\\
$9P_{1/2}$	&	$6S_{1/2}$	&	$1138.63$	&	$1.19[-2]$	&	$1.30[-2]$	&	$8.14(79)[6]$	&	$1.58(15)[-3]$\\
$9P_{1/2}$	&	$6D_{3/2}$	&	$2297.29$	&	$1.12[-1]$	&	$1.04[-1]$	&	$9.33(69)[6]$	&	$3.69(27)[-3]$\\
$9P_{1/2}$	&	$7S_{1/2}$	&	$2409.07$	&	$2.49[-2]$	&	$2.82[-2]$	&	$1.80(24)[6]$	&	$1.57(21)[-3]$\\
$9P_{1/2}$	&	$7D_{3/2}$	&	$5446.84$	&	$1.04[0]$	&	$1.01[0]$	&	$6.51(20)[6]$	&	$1.45(4)[-2]$\\
$9P_{1/2}$	&	$8S_{1/2}$	&	$5772.02$	&	$2.38[-2]$	&	$3.00[-2]$	&	$1.26(31)[5]$	&	 {$6.3(15)[-4]$}\\
$9P_{1/2}$	&	$8D_{3/2}$	&	$27139.20$	&	$2.40[2]$	&	$2.51[2]$	&	$1.21(6)[7]$	&	$6.70(32)[-1]$\\
$9P_{1/2}$	&	$9S_{1/2}$	&	$32636.33$	&	$1.85[2]$	&	$1.89[2]$	&	$5.40(9)[6]$	&	$8.63(14)[-1]$\\
												
$9P_{3/2}$	&	$4D_{3/2}$	&	$431.02$	&	$6.30[-4]$	&	$7.56[-4]$	&	$3.99(76)[6]$	&	$1.11(21)[-4]$\\
$9P_{3/2}$	&	$4D_{5/2}$	&	$433.09$	&	$5.49[-3]$	&	$6.64[-3]$	&	$3.42(68)[7]$	&	 {$6.4(1.3)[-4]$}\\
$9P_{3/2}$	&	$5S_{1/2}$	&	$506.49$	&	$1.21[-2]$	&	$1.27[-2]$	&	$4.72(25)[7]$	&	$3.63(19)[-3]$\\
$9P_{3/2}$	&	$5D_{3/2}$	&	$1073.09$	&	$4.82[-3]$	&	$4.44[-3]$	&	$1.97(16)[6]$	&	$3.41(28)[-4]$\\
$9P_{3/2}$	&	$5D_{5/2}$	&	$1077.00$	&	$4.20[-2]$	&	$3.87[-2]$	&	$1.70(14)[7]$	&	$1.98(16)[-3]$\\
$9P_{3/2}$	&	$6S_{1/2}$	&	$1136.17$	&	$1.53[-2]$	&	$1.73[-2]$	&	$5.28(66)[6]$	&	$2.05(26)[-3]$\\
$9P_{3/2}$	&	$6D_{3/2}$	&	$2287.32$	&	$2.68[-2]$	&	$2.52[-2]$	&	$1.13(7)[6]$	&	$8.90(54)[-4]$\\
$9P_{3/2}$	&	$6D_{5/2}$	&	$2295.78$	&	$2.32[-1]$	&	$2.17[-1]$	&	$9.71(62)[6]$	&	$5.11(33)[-3]$\\
 {$9P_{3/2}$}	&	$7S_{1/2}$	&	$2398.11$	&	$2.27[-2]$	&	$2.74[-2]$	&	 {$8.3(1.6)[5]$}	&	$1.44(28)[-3]$\\
$9P_{3/2}$	&	$7D_{3/2}$	&	$5391.15$	&	$2.50[-1]$	&	$2.45[-1]$	&	$8.08(17)[5]$	&	$3.52(7)[-3]$\\
$9P_{3/2}$	&	$7D_{5/2}$	&	$5417.39$	&	$2.16[0]$	&	$2.11[0]$	&	$6.89(17)[6]$	&	$2.02(5)[-2]$\\
$9P_{3/2}$	&	$8D_{3/2}$	&	$25810.71$	&	$4.61[1]$	&	$4.84[1]$	&	$1.36(7)[6]$	&	$1.36(7)[-1]$\\
$9P_{3/2}$	&	$8D_{5/2}$	&	$26185.44$	&	$4.20[2]$	&	$4.38[2]$	&	$1.19(6)[7]$	&	$8.13(38)[-1]$\\
$9P_{3/2}$	&	$9S_{1/2}$	&	$30734.01$	&	$3.64[2]$	&	$3.70[2]$	&	$6.36(9)[6]$	&	$1.80(3)[0]$\\
												
$9D_{3/2}$	&	$5P_{1/2}$	&	$620.51$	&	$5.17[-3]$	&	$4.72[-3]$	&	$1.10(10)[7]$	&	$1.27(11)[-3]$\\
$9D_{3/2}$	&	$5P_{3/2}$	&	$629.51$	&	$7.08[-4]$	&	$6.35[-4]$	&	$1.44(15)[6]$	&	$8.54(90)[-5]$\\
$9D_{3/2}$	&	$6P_{1/2}$	&	$1306.15$	&	$8.77[-2]$	&	$8.46[-2]$	&	$1.99(7)[7]$	&	$1.02(4)[-2]$\\
$9D_{3/2}$	&	$6P_{3/2}$	&	$1322.55$	&	$1.47[-2]$	&	$1.41[-2]$	&	$3.23(14)[6]$	&	$8.46(36)[-4]$\\
$9D_{3/2}$	&	$7P_{1/2}$	&	$2591.33$	&	$6.40[-1]$	&	$6.32[-1]$	&	$1.86(2)[7]$	&	$3.75(5)[-2]$\\
$9D_{3/2}$	&	$7P_{3/2}$	&	$2624.57$	&	$1.14[-1]$	&	$1.13[-1]$	&	$3.20(5)[6]$	&	$3.31(5)[-3]$\\
$9D_{3/2}$	&	$8P_{1/2}$	&	$5542.67$	&	$5.96[0]$	&	$5.98[0]$	&	$1.77(1)[7]$ &	$1.63(1)[-1]$\\
$9D_{3/2}$	&	$8P_{3/2}$	&	$5632.20$	&	$1.12[0]$	&	$1.13[0]$	&	$3.19(1)[6]$	&	 {$1.516(4)[-2]$}\\
$9D_{3/2}$	&	$9P_{1/2}$	&	$17729.01$	&	$2.51[2]$	&	$2.53[2]$	&	$2.28(2)[7]$	&	$2.15(2)[0]$\\
$9D_{3/2}$	&	$9P_{3/2}$	&	$18345.87$	&	$5.34[1]$	&	$5.38[1]$	&	$4.38(4)[6]$	&	$2.21(2)[-1]$\\
												
$9D_{5/2}$	&	$5P_{3/2}$	&	$629.37$	&	$7.43[-3]$	&	$6.72[-3]$	&	$1.01(10)[7]$	&	$8.97(87)[-4]$\\
$9D_{5/2}$	&	$6P_{3/2}$	&	$1321.90$	&	$1.40[-1]$	&	$1.37[-1]$	&	$2.05(8)[7]$	&	$8.05(32)[-3]$\\
$9D_{5/2}$	&	$7P_{3/2}$	&	$2622.01$	&	$1.07[0]$	&	$1.05[0]$	&	$2.00(3)[7]$	&	$3.09(5)[-2]$\\
$9D_{5/2}$	&	$8P_{3/2}$	&	$5620.43$	&	$1.03[1]$	&	$1.03[1]$	&	 {$1.9615(2)[7]$} &	 {$1.3935(2)[-1]$}\\
$9D_{5/2}$	&	$9P_{3/2}$	&	$18221.68$	&	$4.74[2]$	&	$4.77[2]$	&	$2.65(1)[7]$	&	$1.98(1)[0]$\\
												
$10S_{1/2}$	&	$5P_{1/2}$	&	$618.88$	&	$1.38[-2]$	&	$1.35[-2]$	&	$5.90(13)[7]$	&	$3.39(8)[-3]$\\
$10S_{1/2}$	&	$5P_{3/2}$	&	$627.80$	&	$2.81[-2]$	&	$2.74[-2]$	&	$1.15(3)[8]$	&	$3.40(9)[-3]$\\
$10S_{1/2}$	&	$6P_{1/2}$	&	$1298.78$	&	$6.46[-2]$	&	$6.49[-2]$	&	$2.99(1)[7]$	&	$7.55(4)[-3]$\\
$10S_{1/2}$	&	$6P_{3/2}$	&	$1315.00$	&	$1.31[-1]$	&	$1.31[-1]$	&	 {$5.841(3)[7]$} &	 {$7.572(4)[-3]$}\\
$10S_{1/2}$	&	$7P_{1/2}$	&	$2562.48$	&	$3.18[-1]$	&	$3.25[-1]$	&	$1.92(4)[7]$	&	$1.89(4)[-2]$\\
$10S_{1/2}$	&	$7P_{3/2}$	&	$2594.98$	&	$6.48[-1]$	&	$6.58[-1]$	&	$3.76(6)[7]$	&	$1.90(3)[-2]$\\
$10S_{1/2}$	&	$8P_{1/2}$	&	$5412.36$	&	$2.27[0]$	&	$2.34[0]$	&	$1.45(5)[7]$	&	$6.36(20)[-2]$\\
$10S_{1/2}$	&	$8P_{3/2}$	&	$5497.69$	&	$4.68[0]$	&	$4.78[0]$	&	$2.85(6)[7]$	&	$6.47(14)[-2]$\\
$10S_{1/2}$	&	$9P_{1/2}$	&	$16461.27$	&	$6.06[1]$	&	$6.29[1]$	&	$1.38(5)[7]$	&	$5.59(20)[-1]$\\
$10S_{1/2}$	&	$9P_{3/2}$	&	$16991.74$	&	$1.34[2]$	&	$1.37[2]$	&	$2.76(7)[7]$	&	$5.98(16)[-1]$\\
												
$10P_{1/2}$	&	$4D_{3/2}$	&	$416.88$	&	$1.45[-3]$	&	$1.81[-3]$	&	$2.03(48)[7]$	&	$2.64(63)[-4]$\\
$10P_{1/2}$	&	$5S_{1/2}$	&	$487.08$	&	$4.83[-3]$	&	$5.00[-3]$	&	$4.23(15)[7]$	&	$1.51(5)[-3]$\\
$10P_{1/2}$	&	$5D_{3/2}$	&	$989.54$	&	$1.01[-2]$	&	$9.12[-3]$	&	$1.06(11)[7]$	&	$7.75(77)[-4]$\\
$10P_{1/2}$	&	$6S_{1/2}$	&	$1042.94$	&	$6.91[-3]$	&	$7.53[-3]$	&	$6.17(55)[6]$	&	$1.01(9)[-3]$\\
$10P_{1/2}$	&	$6D_{3/2}$	&	$1938.47$	&	$4.17[-2]$	&	$3.80[-2]$	&	$5.80(53)[6]$	&	$1.64(15)[-3]$\\
$10P_{1/2}$	&	$7S_{1/2}$	&	$2017.46$	&	$1.37[-2]$	&	$1.57[-2]$	&	$1.69(23)[6]$	&	$1.03(14)[-3]$\\
$10P_{1/2}$	&	$7D_{3/2}$	&	$3785.49$	&	$1.95[-1]$	&	$1.84[-1]$	&	$3.64(20)[6]$	&	$3.91(22)[-3]$\\
$10P_{1/2}$	&	$8S_{1/2}$	&	$3939.74$	&	$2.86[-2]$	&	$3.44[-2]$	&	$4.74(91)[5]$	&	$1.10(21)[-3]$\\
$10P_{1/2}$	&	$8D_{3/2}$	&	$8516.33$	&	$1.73[0]$	&	$1.71[0]$	&	$2.84(4)[6]$	&	$1.54(2)[-2]$\\
 {$10P_{1/2}$}	&	$9S_{1/2}$	&	$8991.59$	&	$2.37[-2]$	&	$3.37[-2]$	&	 {$3.3(1.3)[4]$} &	 {$4.0(1.5)[-4]$}\\
$10P_{1/2}$	&	$9D_{3/2}$	&	$41374.25$	&	$4.33[2]$	&	$4.58[2]$	&	$6.19(36)[6]$	&	$7.94(46)[-1]$\\
												
$10P_{3/2}$	&	$4D_{3/2}$	&	$416.63$	&	$4.88[-4]$	&	$5.76[-4]$	&	$3.42(59)[6]$	&	 {$8.9(1.5)[-5]$}\\
$10P_{3/2}$	&	$4D_{5/2}$	&	$418.56$	&	$2.57[-3]$	&	$3.21[-3]$	&	$1.78(42)[7]$	&	$3.11(74)[-4]$\\
$10P_{3/2}$	&	$5S_{1/2}$	&	$486.74$	&	$7.96[-3]$	&	$8.32[-3]$	&	$3.50(16)[7]$	&	$2.48(11)[-3]$\\
$10P_{3/2}$	&	$5D_{3/2}$	&	$988.13$	&	$2.42[-3]$	&	$2.21[-3]$	&	$1.27(11)[6]$	&	$1.86(17)[-4]$\\
$10P_{3/2}$	&	$5D_{5/2}$	&	$991.44$	&	$2.11[-2]$	&	$1.93[-2]$	&	$1.10(10)[7]$	&	$1.08(10)[-3]$\\
$10P_{3/2}$	&	$6S_{1/2}$	&	$1041.37$	&	$9.51[-3]$	&	$1.07[-2]$	&	$4.26(50)[6]$	&	$1.39(16)[-3]$\\
$10P_{3/2}$	&	$6D_{3/2}$	&	$1933.04$	&	$1.02[-2]$	&	$9.49[-3]$	&	$7.18(54)[5]$	&	$4.02(30)[-4]$\\
$10P_{3/2}$	&	$6D_{5/2}$	&	$1939.07$	&	$8.86[-2]$	&	$8.17[-2]$	&	$6.16(49)[6]$	&	$2.31(19)[-3]$\\
 {$10P_{3/2}$}	&	$7S_{1/2}$	&	$2011.58$	&	$1.53[-2]$	&	$1.83[-2]$	&	 {$9.5(1.8)[5]$}	&	$1.15(22)[-3]$\\
$10P_{3/2}$	&	$7D_{3/2}$	&	$3764.81$	&	$4.90[-2]$	&	$4.69[-2]$	&	$4.65(20)[5]$	&	$9.89(43)[-4]$\\
$10P_{3/2}$	&	$7D_{5/2}$	&	$3777.59$	&	$4.21[-1]$	&	$3.99[-1]$	&	$3.96(21)[6]$	&	$5.65(30)[-3]$\\
$10P_{3/2}$	&	$8S_{1/2}$	&	$3917.35$	&	$1.99[-2]$	&	$2.77[-2]$	&	$1.68(60)[5]$	&	 {$7.7(2.8)[-4]$}\\
$10P_{3/2}$	&	$8D_{3/2}$	&	$8412.42$	&	$4.35[-1]$	&	$4.32[-1]$	&	$3.70(2)[5]$	&	$3.92(2)[-3]$\\
$10P_{3/2}$	&	$8D_{5/2}$	&	$8451.84$	&	$3.74[0]$	&	$3.66[0]$	&	$3.14(7)[6]$	&	$2.24(5)[-2]$\\
 {$10P_{3/2}$}	&	$9S_{1/2}$	&	$8875.83$	&	$7.83[-3]$	&	$7.31[-3]$	&	 {$5.3(7.0)[3]$}	&	 {$1.3(1.7)[-4]$}\\
$10P_{3/2}$	&	$9D_{3/2}$	&	$39031.88$	&	$8.32[1]$	&	$8.79[1]$	&	$7.08(40)[5]$	&	$1.62(9)[-1]$\\
$10P_{3/2}$	&	$9D_{5/2}$	&	$39606.22$	&	$7.59[2]$	&	$7.93[2]$	&	$6.18(28)[6]$	&	$9.70(44)[-1]$\\
												
$10D_{3/2}$	&	$5P_{1/2}$	&	$597.21$	&	$2.47[-3]$	&	$2.20[-3]$	&	$5.87(66)[6]$	&	$6.28(71)[-4]$\\
$10D_{3/2}$	&	$5P_{3/2}$	&	$605.50$	&	$3.20[-4]$	&	$2.76[-4]$	&	 {$7.3(1.1)[5]$}	&	$4.02(58)[-5]$\\
$10D_{3/2}$	&	$6P_{1/2}$	&	$1206.84$	&	$4.17[-2]$	&	$3.94[-2]$	&	$1.20(7)[7]$	&	$5.24(29)[-3]$\\
$10D_{3/2}$	&	$6P_{3/2}$	&	$1220.83$	&	$6.86[-3]$	&	$6.42[-3]$	&	$1.91(12)[6]$	&	$4.26(28)[-4]$\\
$10D_{3/2}$	&	$7P_{1/2}$	&	$2227.66$	&	$2.40[-1]$	&	$2.31[-1]$	&	$1.10(4)[7]$	&	$1.63(6)[-2]$\\
$10D_{3/2}$	&	$7P_{3/2}$	&	$2252.18$	&	$4.18[-2]$	&	$4.02[-2]$	&	$1.85(7)[6]$	&	$1.41(5)[-3]$\\
$10D_{3/2}$	&	$8P_{1/2}$	&	$4108.17$	&	$1.32[0]$	&	$1.29[0]$	&	$9.66(21)[6]$	&	$4.89(10)[-2]$\\
$10D_{3/2}$	&	$8P_{3/2}$	&	$4157.14$	&	$2.39[-1]$	&	$2.35[-1]$	&	$1.69(3)[6]$	&	$4.37(8)[-3]$\\
$10D_{3/2}$	&	$9P_{1/2}$	&	$8374.94$	&	$1.07[1]$	&	$1.06[1]$	&	$9.26(11)[6]$	&	$1.95(2)[-1]$\\
$10D_{3/2}$	&	$9P_{3/2}$	&	$8510.11$	&	$2.03[0]$	&	$2.02[0]$	&	$1.67(1)[6]$ &	$1.81(1)[-2]$\\
												
$10D_{5/2}$	&	$5P_{3/2}$	&	$605.40$	&	$3.43[-3]$	&	$3.01[-3]$	&	$5.23(66)[6]$	&	$4.31(54)[-4]$\\
$10D_{5/2}$	&	$6P_{3/2}$	&	$1220.42$	&	$6.54[-2]$	&	$6.15[-2]$	&	$1.22(7)[7]$	&	$4.07(25)[-3]$\\
$10D_{5/2}$	&	$7P_{3/2}$	&	$2250.78$	&	$3.91[-1]$	&	$3.76[-1]$	&	$1.16(4)[7]$	&	$1.32(5)[-2]$\\
$10D_{5/2}$	&	$8P_{3/2}$	&	$4152.37$	&	$2.21[0]$	&	$2.16[0]$	&	$1.04(3)[7]$	&	$4.05(10)[-2]$\\
$10D_{5/2}$	&	$9P_{3/2}$	&	$8490.14$	&	$1.86[1]$	&	$1.83[1]$	&	$1.02(2)[7]$	&	$1.66(3)[-1]$\\
\hline
\end{longtable}

\begin{longtable}{@{\extracolsep\fill}rllllllll@{}}
\caption{\label{tab4} The line strengths ($S_{vk}$) (in a.u.) from both the L and V gauge expressions, wavelengths ($\lambda$) (in $\text{\normalfont\AA}$), transition probabilities ($A_{vk}$) in ($s^{-1}$) and absorption oscillator strengths ($f_{kv}$) for the Nb V ion through E1 decay channel are presented in this table. Values in square brackets represent the order of 10. Uncertainties are given in parentheses.}\\
\hline
Upper State(v) & Lower State(k) & $\lambda$ (in $\text{\normalfont\AA}$) & \multicolumn{2}{c}{$S_{vk}$ (in a.u.)} & $A_{L_{vk}}$(in $s^{-1}$) & $f_{L_{kv}}$\\ 
 & & & L & V & & \\
 \hline
$5P_{1/2}$	&	$4D_{3/2}$	&	$774.02$	&	$1.34[0]$	&	$1.45[0]$	&	$2.94(23)[9]$	&	$1.32(10)[-1]$\\
$5P_{1/2}$	&	$5S_{1/2}$	&	$1877.38$	&	$3.71[0]$	&	$3.88[0]$	&	$5.67(26)[8]$	&	$3.00(14)[-1]$\\
												
$5P_{3/2}$	&	$4D_{3/2}$	&	$753.01$	&	$2.56[-1]$	&	$2.76[-1]$	&	$3.04(23)[8]$	&	$2.58(20)[-2]$\\
$5P_{3/2}$	&	$4D_{5/2}$	&	$763.75$	&	$2.38[0]$	&	$2.57[0]$	&	$2.71(20)[9]$	&	$1.58(12)[-1]$\\
$5P_{3/2}$	&	$5S_{1/2}$	&	$1758.38$	&	$7.43[0]$	&	$7.82[0]$	&	$6.92(36)[8]$	&	$6.42(33)[-1]$\\
												
$5D_{3/2}$	&	$5P_{1/2}$	&	$1212.14$	&	$9.04[0]$	&	$9.33[0]$	&	$2.57(8)[9]$	&	$1.13(4)[0]$\\
$5D_{3/2}$	&	$5P_{3/2}$	&	$1267.52$	&	$1.90[0]$	&	$1.95[0]$	&	$4.72(13)[8]$	&	$1.14(3)[-1]$\\
												
$5D_{5/2}$	&	$5P_{3/2}$	&	$1258.84$	&	$1.70[1]$	&	$1.75[1]$	&	$2.87(9)[9]$	&	$1.02(3)[0]$\\
												
$6S_{1/2}$	&	$5P_{1/2}$	&	$1007.04$	&	$1.28[0]$	&	$1.32[0]$	&	$1.27(4)[9]$	&	$1.92(6)[-1]$\\
$6S_{1/2}$	&	$5P_{3/2}$	&	$1044.97$	&	$2.88[0]$	&	$2.95[0]$	&	$2.55(6)[9]$	&	$2.09(5)[-1]$\\
												
$6P_{1/2}$	&	$4D_{3/2}$	&	$399.19$	&	$5.03[-2]$	&	$5.55[-2]$	&	$8.01(81)[8]$	&	$9.57(96)[-3]$\\
$6P_{1/2}$	&	$5S_{1/2}$	&	$572.81$	&	$5.92[-2]$	&	$6.37[-2]$	&	$3.19(23)[8]$	&	$1.57(12)[-2]$\\
$6P_{1/2}$	&	$5D_{3/2}$	&	$2576.49$	&	$9.64[0]$	&	$9.85[0]$	&	$5.71(12)[8]$	&	$2.84(6)[-1]$\\
$6P_{1/2}$	&	$6S_{1/2}$	&	$4543.35$	&	$1.39[1]$	&	$1.42[1]$	&	$1.50(3)[8]$	&	$4.64(10)[-1]$\\
												
$6P_{3/2}$	&	$4D_{3/2}$	&	$396.79$	&	$1.10[-2]$	&	$1.20[-2]$	&	$8.94(78)[7]$	&	$2.11(19)[-3]$\\
$6P_{3/2}$	&	$4D_{5/2}$	&	$399.75$	&	$1.01[-1]$	&	$1.11[-1]$	&	$8.03(72)[8]$	&	$1.28(12)[-2]$\\
$6P_{3/2}$	&	$5S_{1/2}$	&	$567.88$	&	$8.03[-2]$	&	$8.76[-2]$	&	$2.22(20)[8]$	&	$2.15(19)[-2]$\\
$6P_{3/2}$	&	$5D_{3/2}$	&	$2479.59$	&	$1.83[0]$	&	$1.88[0]$	&	$6.09(15)[7]$	&	$5.61(14)[-2]$\\
$6P_{3/2}$	&	$5D_{5/2}$	&	$2513.52$	&	$1.68[1]$	&	$1.73[1]$	&	$5.37(13)[8]$	&	$3.39(8)[-1]$\\
$6P_{3/2}$	&	$6S_{1/2}$	&	$4250.44$	&	$2.76[1]$	&	$2.82[1]$	&	$1.82(5)[8]$	&	$9.85(25)[-1]$\\
												
$6D_{3/2}$	&	$5P_{1/2}$	&	$633.04$	&	$1.05[-1]$	&	$1.01[-1]$	&	$2.10(9)[8]$	&	$2.52(10)[-2]$\\
$6D_{3/2}$	&	$5P_{3/2}$	&	$647.82$	&	$1.64[-2]$	&	$1.56[-2]$	&	$3.06(16)[7]$	&	$1.92(10)[-3]$\\
$6D_{3/2}$	&	$6P_{1/2}$	&	$2727.98$	&	$2.72[1]$	&	$2.76[1]$	&	$6.78(9)[8]$	&	$1.51(2)[0]$\\
$6D_{3/2}$	&	$6P_{3/2}$	&	$2845.74$	&	$5.72[0]$	&	$5.78[0]$	&	$1.26(1)[8]$	&	$1.53(2)[-1]$\\
												
$6D_{5/2}$	&	$5P_{3/2}$	&	$646.72$	&	$1.62[-1]$	&	$1.54[-1]$	&	$2.02(10)[8]$	&	$1.90(9)[-2]$\\
$6D_{5/2}$	&	$6P_{3/2}$	&	$2824.68$	&	$5.11[1]$	&	$5.16[1]$	&	$7.65(9)[8]$	&	$1.37(2)[0]$\\
												
$7S_{1/2}$	&	$5P_{1/2}$	&	$604.08$	&	$1.29[-1]$	&	$1.29[-1]$	&	$5.94(2)[8]$	&	$3.25(1)[-2]$\\
$7S_{1/2}$	&	$5P_{3/2}$	&	$617.53$	&	$2.72[-1]$	&	$2.70[-1]$	&	$1.17(1)[9]$	&	$3.35(3)[-2]$\\
$7S_{1/2}$	&	$6P_{1/2}$	&	$2260.93$	&	$4.18[0]$	&	$4.33[0]$	&	$3.66(13)[8]$	&	$2.81(10)[-1]$\\
$7S_{1/2}$	&	$6P_{3/2}$	&	$2341.22$	&	$9.35[0]$	&	$9.65[0]$	&	$7.38(23)[8]$	&	$3.03(10)[-1]$\\
												
$7P_{1/2}$	&	$4D_{3/2}$	&	$326.81$	&	$9.14[-3]$	&	$1.23[-2]$	&	$2.65(84)[8]$	&	$2.12(68)[-3]$\\
$7P_{1/2}$	&	$5S_{1/2}$	&	$434.68$	&	$2.83[-2]$	&	$2.93[-2]$	&	$3.49(13)[8]$	&	$9.87(36)[-3]$\\
$7P_{1/2}$	&	$5D_{3/2}$	&	$1060.53$	&	$2.23[-1]$	&	$2.10[-1]$	&	$1.89(11)[8]$	&	$1.59(9)[-2]$\\
$7P_{1/2}$	&	$6S_{1/2}$	&	$1290.49$	&	$6.01[-2]$	&	$6.62[-2]$	&	$2.83(28)[7]$	&	$7.08(69)[-3]$\\
$7P_{1/2}$	&	$6D_{3/2}$	&	$5312.68$	&	$3.02[1]$	&	$3.10[1]$	&	$2.04(5)[8]$	&	$4.31(11)[-1]$\\
$7P_{1/2}$	&	$7S_{1/2}$	&	$8888.49$	&	$3.46[1]$	&	$3.50[1]$	&	$5.00(5)[7]$	&	$5.92(6)[-1]$\\
												
 {$7P_{3/2}$}	&	$4D_{3/2}$	&	$325.96$	&	$2.27[-3]$	&	$2.78[-3]$	&	$3.31(72)[7]$	&	 {$5.3(1.2)[-4]$}\\
$7P_{3/2}$	&	$4D_{5/2}$	&	$327.95$	&	$2.01[-2]$	&	$2.48[-2]$	&	$2.88(64)[8]$	&	$3.10(69)[-3]$\\
$7P_{3/2}$	&	$5S_{1/2}$	&	$433.17$	&	$4.38[-2]$	&	$4.58[-2]$	&	$2.73(13)[8]$	&	$1.54(7)[-2]$\\
$7P_{3/2}$	&	$5D_{3/2}$	&	$1051.59$	&	$4.86[-2]$	&	$4.62[-2]$	&	$2.12(11)[7]$	&	$3.51(18)[-3]$\\
$7P_{3/2}$	&	$5D_{5/2}$	&	$1057.64$	&	$4.32[-1]$	&	$4.10[-1]$	&	$1.85(10)[8]$	&	$2.07(11)[-2]$\\
 {$7P_{3/2}$}	&	$6S_{1/2}$	&	$1277.27$	&	$6.00[-2]$	&	$6.88[-2]$	&	$1.46(21)[7]$	&	 {$7.1(1.0)[-3]$}\\
$7P_{3/2}$	&	$6D_{3/2}$	&	$5095.67$	&	$5.74[0]$	&	$5.92[0]$	&	$2.20(7)[7]$	&	$8.56(26)[-2]$\\
$7P_{3/2}$	&	$6D_{5/2}$	&	$5164.62$	&	$5.26[1]$	&	$5.42[1]$	&	$1.93(6)[8]$	&	$5.16(15)[-1]$\\
$7P_{3/2}$	&	$7S_{1/2}$	&	$8297.31$	&	$6.85[1]$	&	$6.94[1]$	&	$6.07(8)[7]$	&	$1.25(2)[0]$\\
												
$7D_{3/2}$	&	$5P_{1/2}$	&	$508.88$	&	$7.89[-3]$	&	$6.74[-3]$	&	$3.03(46)[7]$	&	$2.35(36)[-3]$\\
$7D_{3/2}$	&	$5P_{3/2}$	&	$518.39$	&	$8.58[-4]$	&	$6.86[-4]$	&	$3.12(66)[6]$	&	$1.26(27)[-4]$\\
$7D_{3/2}$	&	$6P_{1/2}$	&	$1329.80$	&	$5.85[-1]$	&	$5.69[-2]$	&	$1.26(3)[8]$	&	$6.68(19)[-2]$\\
$7D_{3/2}$	&	$6P_{3/2}$	&	$1357.18$	&	$1.00[-1]$	&	$9.72[-2]$	&	$2.03(7)[7]$	&	$5.62(18)[-3]$\\
$7D_{3/2}$	&	$7P_{1/2}$	&	$5071.20$	&	$6.03[1]$	&	$6.08[1]$	&	$2.34(2)[8]$	&	$1.81(1)[0]$\\
$7D_{3/2}$	&	$7P_{3/2}$	&	$5286.08$	&	$1.27[1]$	&	$1.28[1]$	&	$4.37(2)[7]$	&	$1.83(1)[-1]$\\
												
$7D_{5/2}$	&	$5P_{3/2}$	&	$517.99$	&	$9.96[-3]$	&	$8.22[-3]$	&	$2.42(44)[7]$	&	$1.46(27)[-3]$\\
$7D_{5/2}$	&	$6P_{3/2}$	&	$1354.44$	&	$9.49[-1]$	&	$9.20[-1]$	&	$1.29(4)[8]$	&	$5.32(17)[-2]$\\
$7D_{5/2}$	&	$7P_{3/2}$	&	$5244.75$	&	$1.13[2]$	&	$1.14[2]$	&	$2.66(1)[8]$	&	$1.64(1)[0]$\\
												
$8S_{1/2}$	&	$5P_{1/2}$	&	$498.28$	&	$4.13[-2]$	&	$4.06[-2]$	&	$3.38(6)[8]$	&	$1.26(2)[-2]$\\
$8S_{1/2}$	&	$5P_{3/2}$	&	$507.40$	&	$8.54[-2]$	&	$8.35[-2]$	&	$6.63(15)[8]$	&	$1.28(3)[-2]$\\
$8S_{1/2}$	&	$6P_{1/2}$	&	$1259.80$	&	$3.57[-1]$	&	$3.63[-1]$	&	$1.81(3)[8]$	&	$4.31(6)[-2]$\\
$8S_{1/2}$	&	$6P_{3/2}$	&	$1284.34$	&	$7.50[-1]$	&	$7.57[-1]$	&	$3.59(3)[8]$	&	$4.43(4)[-2]$\\
$8S_{1/2}$	&	$7P_{1/2}$	&	$4184.50$	&	$1.02[1]$	&	$1.06[1]$	&	$1.41(5)[8]$	&	$3.71(14)[-1]$\\
$8S_{1/2}$	&	$7P_{3/2}$	&	$4329.74$	&	$2.28[1]$	&	$2.36[1]$	&	$2.84(9)[8]$	&	$4.00(13)[-1]$\\
												
 {$8P_{1/2}$}	&	$4D_{3/2}$	&	$307.43$	&	$2.44[-3]$	&	$4.19[-3]$	&	 {$8.5(5.3)[7]$}	&	 {$6.0(3.7)[-4]$}\\
$8P_{1/2}$	&	$5S_{1/2}$	&	$394.99$	&	$1.69[-2]$	&	$1.72[-2]$	&	$2.77(5)[8]$	&	$6.49(12)[-3]$\\
$8P_{1/2}$	&	$5D_{3/2}$	&	$823.88$	&	$5.21[-2]$	&	$4.76[-2]$	&	$9.44(84)[7]$	&	$4.80(43)[-3]$\\
$8P_{1/2}$	&	$6S_{1/2}$	&	$944.92$	&	$2.86[-2]$	&	$3.06[-2]$	&	$3.43(24)[7]$	&	$4.60(32)[-3]$\\
$8P_{1/2}$	&	$6D_{3/2}$	&	$2082.48$	&	$5.58[-1]$	&	$5.32[-1]$	&	$6.26(29)[7]$	&	$2.03(9)[-2]$\\
$8P_{1/2}$	&	$7S_{1/2}$	&	$2433.04$	&	$7.52[-2]$	&	$8.23[-2]$	&	$5.29(48)[6]$	&	$4.70(43)[-3]$\\
$8P_{1/2}$	&	$7D_{3/2}$	&	$9672.58$	&	$7.10[1]$	&	$7.34[1]$	&	$7.95(26)[7]$	&	$5.58(18)[-1]$\\
$8P_{1/2}$	&	$8S_{1/2}$	&	$15199.75$	&	$7.16[1]$	&	$7.17[1]$	&	 {$2.065(4)[7]$}	&	$7.15(1)[-1]$\\
												
$8P_{3/2}$	&	$4D_{3/2}$	&	$306.99$	&	$6.97[-4]$	&	$1.01[-3]$	&	$1.22(50)[7]$	&	$1.72(71)[-4]$\\
$8P_{3/2}$	&	$4D_{5/2}$	&	$308.64$	&	$6.12[-3]$	&	$8.97[-3]$	&	$1.05(45)[8]$	&	$1.00(42)[-3]$\\
$8P_{3/2}$	&	$5S_{1/2}$	&	$394.27$	&	$2.65[-2]$	&	$2.73[-2]$	&	$2.19(6)[8]$	&	$1.02(3)[-2]$\\
$8P_{3/2}$	&	$5D_{3/2}$	&	$820.74$	&	$1.16[-2]$	&	$1.07[-2]$	&	$1.06(8)[7]$	&	$1.08(9)[-3]$\\
$8P_{3/2}$	&	$5D_{5/2}$	&	$824.33$	&	$1.03[-1]$	&	$9.47[-2]$	&	$9.29(75)[7]$	&	$6.31(51)[-3]$\\
$8P_{3/2}$	&	$6S_{1/2}$	&	$940.79$	&	$3.72[-2]$	&	$4.08[-2]$	&	$2.27(21)[7]$	&	$6.01(56)[-3]$\\
$8P_{3/2}$	&	$6D_{3/2}$	&	$2062.53$	&	$1.24[-1]$	&	$1.19[-1]$	&	$7.15(28)[6]$	&	$4.56(18)[-3]$\\
$8P_{3/2}$	&	$6D_{5/2}$	&	$2073.40$	&	$1.09[0]$	&	$1.05[0]$	&	$6.21(25)[7]$	&	$2.67(11)[-2]$\\
$8P_{3/2}$	&	$7S_{1/2}$	&	$2405.86$	&	$5.58[-2]$	&	$6.53[-2]$	&	$2.03(33)[6]$	&	$3.52(58)[-3]$\\
$8P_{3/2}$	&	$7D_{3/2}$	&	$9256.70$	&	$1.35[1]$	&	$1.40[1]$	&	$8.62(30)[6]$	&	$1.11(4)[-1]$\\
$8P_{3/2}$	&	$7D_{5/2}$	&	$9380.50$	&	$1.24[2]$	&	$1.28[2]$	&	$7.59(25)[7]$	&	$6.67(22)[-1]$\\
$8P_{3/2}$	&	$8S_{1/2}$	&	$14197.42$	&	$1.41[2]$	&	$1.41[2]$	&	$2.50(1)[7]$	&	 {$1.510(4)[0]$}\\
												
 {$8D_{3/2}$}	&	$5P_{1/2}$	&	$468.43$	&	$7.18[-4]$	&	$4.58[-4]$	&	 {$3.5(1.4)[6]$}	&	$2.33(94)[-4]$\\
 {$8D_{3/2}$}	&	$5P_{3/2}$	&	$476.00$	&	$2.50[-5]$	&	$6.25[-6]$	&	 {$1.2(1.2)[5]$}	&	 {$4.0(4.0)[-6]$}\\
$8D_{3/2}$	&	$6P_{1/2}$	&	$1042.70$	&	$9.38[-2]$	&	$8.78[-2]$	&	$4.19(27)[7]$	&	$1.37(9)[-2]$\\
$8D_{3/2}$	&	$6P_{3/2}$	&	$1058.70$	&	$1.45[-2]$	&	$1.34[-2]$	&	$6.20(48)[6]$	&	$1.04(8)[-3]$\\
$8D_{3/2}$	&	$7P_{1/2}$	&	$2401.87$	&	$1.54[0]$	&	$1.52[0]$	&	$5.62(8)[7]$	&	$9.72(13)[-2]$\\
$8D_{3/2}$	&	$7P_{3/2}$	&	$2446.65$	&	$2.72[-1]$	&	$2.67[-1]$	&	$9.42(19)[6]$	&	$8.45(17)[-3]$\\
$8D_{3/2}$	&	$8P_{1/2}$	&	$8491.33$	&	$1.15[2]$	&	$1.16[2]$	&	$9.52(9)[7]$	&	$2.06(2)[0]$\\
$8D_{3/2}$	&	$8P_{3/2}$	&	$8839.98$	&	$2.44[1]$	&	$2.45[1]$	&	$1.79(1)[7]$	&	$2.10(1)[-1]$\\
												
 {$8D_{5/2}$}	&	$5P_{3/2}$	&	$475.80$	&	$5.76[-4]$	&	$2.86[-4]$	&	 {$1.8(1.1)[6]$}	&	 {$9.2(5.4)[-5]$}\\
$8D_{5/2}$	&	$6P_{3/2}$	&	$1057.71$	&	$1.42[-1]$	&	$1.31[-1]$	&	$4.04(30)[7]$	&	$1.02(7)[-2]$\\
$8D_{5/2}$	&	$7P_{3/2}$	&	$2441.38$	&	$2.55[0]$	&	$2.50[0]$	&	$5.91(12)[7]$	&	$7.93(16)[-2]$\\
 {$8D_{5/2}$}	&	$8P_{3/2}$	&	$8771.50$	&	$2.17[2]$	&	$2.18[2]$	&	 {$1.1(4.4)[8]$}	&	 {$1.9(7.5)[0]$}\\
												
$9S_{1/2}$	&	$5P_{1/2}$	&	$463.54$	&	$1.92[-2]$	&	$1.88[-2]$	&	$1.96(5)[8]$	&	$6.30(16)[-3]$\\
$9S_{1/2}$	&	$5P_{3/2}$	&	$470.95$	&	$3.94[-2]$	&	$3.82[-2]$	&	$3.83(12)[8]$	&	$6.36(20)[-3]$\\
$9S_{1/2}$	&	$6P_{1/2}$	&	$1018.77$	&	$1.05[-1]$	&	$1.06[-1]$	&	$1.01(1)[8]$	&	$1.57(1)[-2]$\\
$9S_{1/2}$	&	$6P_{3/2}$	&	$1034.03$	&	$2.18[-1]$	&	$2.18[-1]$	&	 {$1.997(3)[8]$} &	 {$1.601(2)[-2]$}\\
$9S_{1/2}$	&	$7P_{1/2}$	&	$2278.56$	&	$8.00[-1]$	&	$8.20[-1]$	&	$6.85(18)[7]$	&	$5.33(14)[-2]$\\
$9S_{1/2}$	&	$7P_{3/2}$	&	$2318.83$	&	$1.68[0]$	&	$1.70[0]$	&	$1.36(2)[8]$	&	$5.49(8)[-2]$\\
$9S_{1/2}$	&	$8P_{1/2}$	&	$7127.67$	&	$2.11[1]$	&	$2.21[1]$	&	$5.91(26)[7]$	&	$4.51(20)[-1]$\\
$9S_{1/2}$	&	$8P_{3/2}$	&	$7371.72$	&	$4.73[1]$	&	$4.88[1]$	&	$1.20(4)[8]$	&	$4.88(15)[-1]$\\
												
 {$9P_{1/2}$}	&	$4D_{3/2}$	&	$291.12$	&	$6.86[-4]$	&	$1.73[-3]$	&	 {$2.8(3.3)[7]$}	&	 {$1.8(2.1)[-4]$}\\
$9P_{1/2}$	&	$5S_{1/2}$	&	$368.48$	&	$1.91[-2]$	&	$1.91[-2]$	&	$3.86(1)[8]$	&	$7.86(1)[-3]$\\
$9P_{1/2}$	&	$5D_{3/2}$	&	$716.36$	&	$2.14[-2]$	&	$1.92[-2]$	&	$5.89(62)[7]$	&	$2.27(24)[-3]$\\
$9P_{1/2}$	&	$6S_{1/2}$	&	$806.14$	&	$1.53[-2]$	&	$1.61[-2]$	&	$2.95(16)[7]$	&	$2.88(16)[-3]$\\
$9P_{1/2}$	&	$6D_{3/2}$	&	$1509.72$	&	$1.25[-1]$	&	$1.16[-1]$	&	$3.67(27)[7]$	&	$6.27(46)[-3]$\\
$9P_{1/2}$	&	$7S_{1/2}$	&	$1685.81$	&	$3.97[-2]$	&	$4.24[-2]$	&	$8.40(57)[6]$	&	$3.58(24)[-3]$\\
$9P_{1/2}$	&	$7D_{3/2}$	&	$3501.83$	&	$1.14[0]$	&	$1.12[0]$	&	$2.69(5)[7]$	&	$2.47(5)[-2]$\\
$9P_{1/2}$	&	$8S_{1/2}$	&	$4032.74$	&	$1.04[-1]$	&	$1.10[-1]$	&	$1.61(9)[6]$	&	$3.92(22)[-3]$\\
$9P_{1/2}$	&	$8D_{3/2}$	&	$15524.92$	&	$1.43[2]$	&	$1.49[2]$	&	$3.86(17)[7]$	&	$6.98(31)[-1]$\\
$9P_{1/2}$	&	$9S_{1/2}$	&	$23876.94$	&	$1.31[2]$	&	$1.31[2]$	&	$9.74(2)[6]$	&	$8.32(1)[-1]$\\
												
 {$9P_{3/2}$}	&	$4D_{3/2}$	&	$290.83$	&	$2.46[-4]$	&	$4.49[-4]$	&	 {$5.1(3.6)[6]$}	&	 {$6.4(4.5)[-5]$}\\
 {$9P_{3/2}$}	&	$4D_{5/2}$	&	$292.31$	&	$2.13[-3]$	&	$3.96[-3]$	&	 {$4.3(3.1)[7]$}	&	 {$3.7(2.7)[-4]$}\\
$9P_{3/2}$	&	$5S_{1/2}$	&	$368.00$	&	$1.33[-2]$	&	$1.35[-2]$	&	$1.35(3)[8]$	&	$5.48(11)[-3]$\\
$9P_{3/2}$	&	$5D_{3/2}$	&	$714.57$	&	$4.82[-3]$	&	$4.37[-3[$	&	$6.69(64)[6]$	&	$5.12(49)[-4]$\\
$9P_{3/2}$	&	$5D_{5/2}$	&	$717.29$	&	$4.25[-2]$	&	$3.85[-2]$	&	$5.84(57)[7]$	&	$3.00(29)[-3]$\\
$9P_{3/2}$	&	$6S_{1/2}$	&	$803.88$	&	$2.14[-2]$	&	$2.30[-2]$	&	$2.08(16)[7]$	&	$4.04(30)[-3]$\\
$9P_{3/2}$	&	$6D_{3/2}$	&	$1501.79$	&	$2.83[-2]$	&	$2.65[-2]$	&	$4.24(28)[6]$	&	$1.43(9)[-3]$\\
$9P_{3/2}$	&	$6D_{5/2}$	&	$1507.54$	&	$2.49[-1]$	&	$2.32[-1]$	&	$3.67(25)[7]$	&	$8.35(57)[-3]$\\
$9P_{3/2}$	&	$7S_{1/2}$	&	$1675.92$	&	$4.47[-2]$	&	$4.98[-2]$	&	$4.81(53)[6]$	&	$4.05(45)[-3]$\\
$9P_{3/2}$	&	$7D_{3/2}$	&	$3459.45$	&	$2.59[-1]$	&	$2.54[-1]$	&	$3.17(7)[6]$	&	$5.69(12)[-3]$\\
$9P_{3/2}$	&	$7D_{5/2}$	&	$3476.60$	&	$2.28[0]$	&	$2.22[0]$	&	$2.74(7)[7]$	&	$3.31(9)[-2]$\\
 {$9P_{3/2}$}	&	$8S_{1/2}$	&	$3976.63$	&	$5.18[-2]$	&	$6.40[-2]$	&	$4.18(92)[5]$	&	$1.98(44)[-3]$\\
$9P_{3/2}$	&	$8D_{3/2}$	&	$14725.11$	&	$2.70[1]$	&	$2.80[1]$	&	$4.29(16)[6]$	&	$1.39(5)[-1]$\\
$9P_{3/2}$	&	$8D_{5/2}$	&	$14919.13$	&	$2.47[2]$	&	$2.55[2]$	&	$3.77(13)[7]$	&	$8.38(28)[-1]$\\
$9P_{3/2}$	&	$9S_{1/2}$	&	$22036.11$	&	$2.57[2]$	&	$2.53[2]$	&	$1.22(2)[7]$	&	$1.77(3)[0]$\\
												
 {$9D_{3/2}$}	&	$5P_{3/2}$	&	$446.02$	&	$1.16[-5]$	&	$2.81[-5]$	&	 {$6.6(7.4)[4]$}	&	 {$2.0(2.2)[-6]$}\\
$9D_{3/2}$	&	$6P_{1/2}$	&	$908.87$	&	$2.82[-2]$	&	$2.56[-2]$	&	$1.90(18)[7]$	&	$4.71(44)[-3]$\\
$9D_{3/2}$	&	$6P_{3/2}$	&	$921.00$	&	$4.03[-3]$	&	$3.59[-3]$	&	$2.61(30)[6]$	&	$3.32(38)[-4]$\\
$9D_{3/2}$	&	$7P_{1/2}$	&	$1793.52$	&	$2.75[-1]$	&	$2.68[-1]$	&	$2.42(6)[7]$	&	$2.33(6)[-2]$\\
$9D_{3/2}$	&	$7P_{3/2}$	&	$1818.37$	&	$4.56[-2]$	&	$4.38[-2]$	&	$3.84(16)[6]$	&	$1.91(8)[-3]$\\
$9D_{3/2}$	&	$8P_{1/2}$	&	$3861.18$	&	$3.13[0]$	&	$3.18[0]$	&	$2.76(4)[7]$	&	$1.23(2)[-1]$\\
$9D_{3/2}$	&	$8P_{3/2}$	&	$3931.69$	&	$5.67[-1]$	&	$5.65[0]$	&	$4.73(2)[6]$	&	 {$1.096(4)[-2]$}\\
$9D_{3/2}$	&	$9P_{1/2}$	&	$13019.37$	&	$1.96[2]$	&	$2.02[2]$	&	$4.50(15)[7]$	&	$2.29(7)[0]$\\
$9D_{3/2}$	&	$9P_{3/2}$	&	$13640.71$	&	$4.25[1]$	&	$4.30[1]$	&	$8.47(11)[6]$	&	$2.36(3)[-1]$\\
												
$9D_{5/2}$	&	$6P_{3/2}$	&	$920.50$	&	$4.04[-2]$	&	$3.61[-2]$	&	$1.75(19)[7]$	&	$3.33(37)[-3]$\\
$9D_{5/2}$	&	$7P_{3/2}$	&	$1816.42$	&	$4.36[-1]$	&	$4.17[-1]$	&	$2.46(11)[7]$	&	$1.82(8)[-2]$\\
$9D_{5/2}$	&	$8P_{3/2}$	&	$3922.57$	&	$5.28[0]$	&	$5.23[0]$	&	$2.96(3)[7]$	&	$1.02(1)[-1]$\\
$9D_{5/2}$	&	$9P_{3/2}$	&	$13531.61$	&	$3.78[2]$	&	$3.81[2]$	&	$5.15(5)[7]$	&	$2.12(2)[0]$\\
												
$10S_{1/2}$	&	$5P_{1/2}$	&	$436.74$	&	$1.06[-2]$	&	$1.02[-2]$	&	$1.29(4)[8]$	&	$3.68(11)[-3]$\\
$10S_{1/2}$	&	$5P_{3/2}$	&	$443.30$	&	$2.15[-2]$	&	$2.08[-2]$	&	$2.50(9)[8]$	&	$3.69(13)[-3]$\\
$10S_{1/2}$	&	$6P_{1/2}$	&	$897.68$	&	$4.59[-2]$	&	$4.61[-2]$	&	$6.43(2)[7]$	&	$7.77(3)[-3]$\\
$10S_{1/2}$	&	$6P_{3/2}$	&	$909.51$	&	$9.45[-2]$	&	$9.38[-2]$	&	$1.27(1)[8]$	&	$7.89(6)[-3]$\\
$10S_{1/2}$	&	$7P_{1/2}$	&	$1750.46$	&	$2.20[-1]$	&	$2.26[-1]$	&	$4.16(11)[7]$	&	$1.91(5)[-2]$\\
$10S_{1/2}$	&	$7P_{3/2}$	&	$1774.13$	&	$4.60[-1]$	&	$4.60[-1]$	&	$8.34(0)[7]$	&	$1.97(0)[-2]$\\
$10S_{1/2}$	&	$8P_{1/2}$	&	$3666.98$	&	$1.53[0]$	&	$1.60[0]$	&	$3.14(14)[7]$	&	$6.33(27)[-2]$\\
$10S_{1/2}$	&	$8P_{3/2}$	&	$3730.52$	&	$3.26[0]$	&	$3.26[0]$	&	$6.35(1)[7]$	&	$6.63(1)[-2]$\\
$10S_{1/2}$	&	$9P_{1/2}$	&	$11046.79$	&	$3.87]1]$	&	$4.13[1]$	&	$2.91(19)[7]$	&	$5.33(35)[-1]$\\
$10S_{1/2}$	&	$9P_{3/2}$	&	$11490.90$	&	$8.96[1]$	&	$9.15[1]$	&	$5.98(12)[7]$	&	$5.92(12)[-1]$\\
												
$10P_{1/2}$	&	$5S_{1/2}$	&	$353.07$	&	$7.48[-3]$	&	$5.18[-3]$	&	$1.72(58)[8]$	&	$3.2(11)[-3]$\\
$10P_{1/2}$	&	$5D_{3/2}$	&	$660.34$	&	$1.16[-2]$	&	$1.03[-2]$	&	$4.07(46)[7]$	&	$1.33(15)[-3]$\\
$10P_{1/2}$	&	$6S_{1/2}$	&	$735.89$	&	$9.60[-3]$	&	$1.01[-2]$	&	$2.44(13)[7]$	&	$1.98(11)[-3]$\\
$10P_{1/2}$	&	$6D_{3/2}$	&	$1280.73$	&	$5.13[-2]$	&	$4.72[-2]$	&	$2.48(20)[7]$	&	$3.04(25)[-3]$\\
$10P_{1/2}$	&	$7S_{1/2}$	&	$1405.25$	&	$2.21[-2]$	&	$2.37[-2]$	&	$8.08(54)[6]$	&	$2.39(16)[-3]$\\
$10P_{1/2}$	&	$7D_{3/2}$	&	$2475.27$	&	$2.42[-1]$	&	$2.39[-1]$	&	$1.61(2)[7]$	&	$7.42(9)[-3]$\\
$10P_{1/2}$	&	$8S_{1/2}$	&	$2729.25$	&	$6.29[-2]$	&	$6.54[-2]$	&	$3.13(12)[6]$	&	$3.50(14)[-3]$\\
$10P_{1/2}$	&	$8D_{3/2}$	&	$5469.16$	&	$1.94[0]$	&	$2.09[0]$	&	$1.20(9)[7]$	&	$2.70(20)[-2]$\\
$10P_{1/2}$	&	$9S_{1/2}$	&	$6237.82$	&	$1.86[-1]$	&	$1.73[-1]$	&	$7.76(54)[5]$	&	$4.53(32)[-3]$\\
$10P_{1/2}$	&	$9D_{3/2}$	&	$24025.60$	&	$2.62[2]$	&	$2.88[2]$	&	$1.91(19)[7]$	&	$8.27(82)[-1]$\\
												
 {$10P_{3/2}$}	&	$4D_{3/2}$	&	$281.10$	&	$9.22[-5]$	&	$2.28[-4]$	&	 {$2.1(2.4)[6]$}	&	 {$2.5(2.9)[-5]$}\\
 {$10P_{3/2}$}	&	$4D_{5/2}$	&	$282.48$	&	$7.84[-4]$	&	$2.02[-3]$	&	 {$1.8(2.1)[7]$}	&	 {$1.4(1.7)[-4]$}\\
 {$10P_{3/2}$}	&	$5S_{1/2}$	&	$352.57$	&	$1.30[-2]$	&	$8.76[-3]$	&	$1.50(54)[8]$	&	 {$5.6(2.0)[-3]$}\\
$10P_{3/2}$	&	$5D_{3/2}$	&	$658.59$	&	$2.64[-3]$	&	$2.37[-3]$	&	$4.68(49)[6]$	&	$3.05(32)[-4]$\\
$10P_{3/2}$	&	$5D_{5/2}$	&	$660.90$	&	$2.33[-2]$	&	$2.09[-2]$	&	$4.10(44)[7]$	&	$1.79(19)[-3]$\\
$10P_{3/2}$	&	$6S_{1/2}$	&	$733.72$	&	$1.39[-2]$	&	$1.50[-2]$	&	$1.79(13)[7]$	&	$2.89(21)[-3]$\\
$10P_{3/2}$	&	$6D_{3/2}$	&	$1274.17$	&	$1.19[-2]$	&	$1.10[-2]$	&	$2.91(22)[6]$	&	$7.09(55)[-4]$\\
$10P_{3/2}$	&	$6D_{5/2}$	&	$1278.31$	&	$1.04[-1]$	&	$9.58[-2]$	&	$2.53(21)[7]$	&	$4.13(35)[-3]$\\
$10P_{3/2}$	&	$7S_{1/2}$	&	$1397.36$	&	$2.73[-2]$	&	$3.05[-2]$	&	$5.06(58)[6]$	&	$2.96(34)[-3]$\\
$10P_{3/2}$	&	$7D_{3/2}$	&	$2450.91$	&	$5.83[-2]$	&	$5.62[-2]$	&	$2.00(7)[6]$	&	$1.81(6)[-3]$\\
$10P_{3/2}$	&	$7D_{5/2}$	&	$2459.51$	&	$5.09[-1]$	&	$4.86[-1]$	&	$1.73(8)[7]$	&	$1.05(5)[-2]$\\
$10P_{3/2}$	&	$8S_{1/2}$	&	$2699.66$	&	$5.66[-2]$	&	$6.68[-2]$	&	$1.46(25)[6]$	&	$3.18(55)[-3]$\\
$10P_{3/2}$	&	$8D_{3/2}$	&	$5351.62$	&	$4.86[-1]$	&	$4.91[-1]$	&	$1.61(2)[6]$	&	$6.90(7)[-3]$\\
$10P_{3/2}$	&	$8D_{5/2}$	&	$5377.03$	&	$4.26[0]$	&	$4.23[0]$	&	$1.39(1)[7]$	&	$4.01(2)[-2]$\\
 {$10P_{3/2}$}	&	$9S_{1/2}$	&	$6085.38$	&	$3.29[-2]$	&	$5.75[-2]$	&	 {$7.4(4.8)[4]$}	&	 {$8.2(5.3)[-4]$}\\
$10P_{3/2}$	&	$9D_{3/2}$	&	$21911.52$	&	$4.86[1]$	&	$5.11[1]$	&	$2.34(12)[6]$	&	$1.69(9)[-1]$\\
$10P_{3/2}$	&	$9D_{5/2}$	&	$22198.99$	&	$4.45[2]$	&	$4.61[2]$	&	$2.06(7)[7]$	&	$1.01(4)[0]$\\
												
 {$10D_{3/2}$}	&	$5P_{1/2}$	&	$421.63$	&	$3.36[-5]$	&	$9.03[-5]$	&	 {$2.3(2.9)[5]$}	&	 {$1.2(1.6)[-5]$}\\
 {$10D_{3/2}$}	&	$5P_{3/2}$	&	$427.75$	&	$4.36[-5]$	&	$6.72[-5]$	&	 {$2.8(1.4)[5]$}	&	 {$7.7(3.8)[-6]$}\\
 {$10D_{3/2}$}	&	$6P_{1/2}$	&	$836.10$	&	$1.15[-2]$	&	$1.03[-2]$	&	 {$10.0(1.1)[6]$}	&	$2.09(23)[-3]$\\
$10D_{3/2}$	&	$6P_{3/2}$	&	$846.35$	&	$1.55[-3]$	&	$1.33[-3]$	&	$1.30(19)[6]$	&	$1.39(21)[-4]$\\
$10D_{3/2}$	&	$7P_{1/2}$	&	$1530.62$	&	$8.92[-2]$	&	$8.76[-2]$	&	$1.26(2)[7]$	&	$8.85(15)[-3]$\\
$10D_{3/2}$	&	$7P_{3/2}$	&	$1548.69$	&	$1.42[-2]$	&	$1.36[-2]$	&	$1.93(8)[6]$	&	$6.96(28)[-4]$\\
$10D_{3/2}$	&	$8P_{1/2}$	&	$2818.87$	&	$5.67[-1]$	&	$5.99[-1]$	&	$1.28(7)[7]$ &	$3.06(17)[-2]$\\
$10D_{3/2}$	&	$8P_{3/2}$	&	$2856.26$	&	$9.80[-2]$	&	$9.98[-2]$	&	$2.13(4)[6]$	&	$2.60(5)[-3]$\\
$10D_{3/2}$	&	$9P_{1/2}$	&	$5794.65$	&	$5.43[0]$	&	$6.08[0]$	&	$1.41(17)[7]$	&	$1.42(17)[-1]$\\
$10D_{3/2}$	&	$9P_{3/2}$	&	$5914.56$	&	$9.98[-1]$	&	$1.06[0]$	&	$2.44(15)[6]$	&	$1.28(8)[-2]$\\
										
 {$10D_{5/2}$}	&	$5P_{3/2}$	&	$427.67$	&	$1.64[-4]$	&	$3.06[-4]$	&	 {$7.1(5.2)[5]$}	&	 {$2.9(2.1)[-5]$}\\
 {$10D_{5/2}$}	&	$6P_{3/2}$	&	$846.04$	&	$1.58[-2]$	&	$1.37[-2]$	&	 {$8.8(1.2)[6]$}	&	$1.42(20)[-3]$\\
$10D_{5/2}$	&	$7P_{3/2}$	&	$1547.66$	&	$1.38[-1]$	&	$1.31[-1]$	&	$1.25(6)[7]$	&	$6.75(34)[-3]$\\
$10D_{5/2}$	&	$8P_{3/2}$	&	$2852.75$	&	$9.29[-1]$	&	$9.27[-1]$	&	 {$1.352(3)[7]$}	&	$2.47(1)[-2]$\\
$10D_{5/2}$	&	$9P_{3/2}$	&	$5899.51$	&	$9.28[0]$	&	$9.59[0]$	&	$1.53(5)[7]$	&	$1.19(4)[-1]$\\
\hline
\end{longtable}
\clearpage
\newpage

\begin{figure}
\centering
\includegraphics[width=\textwidth,height=20cm]{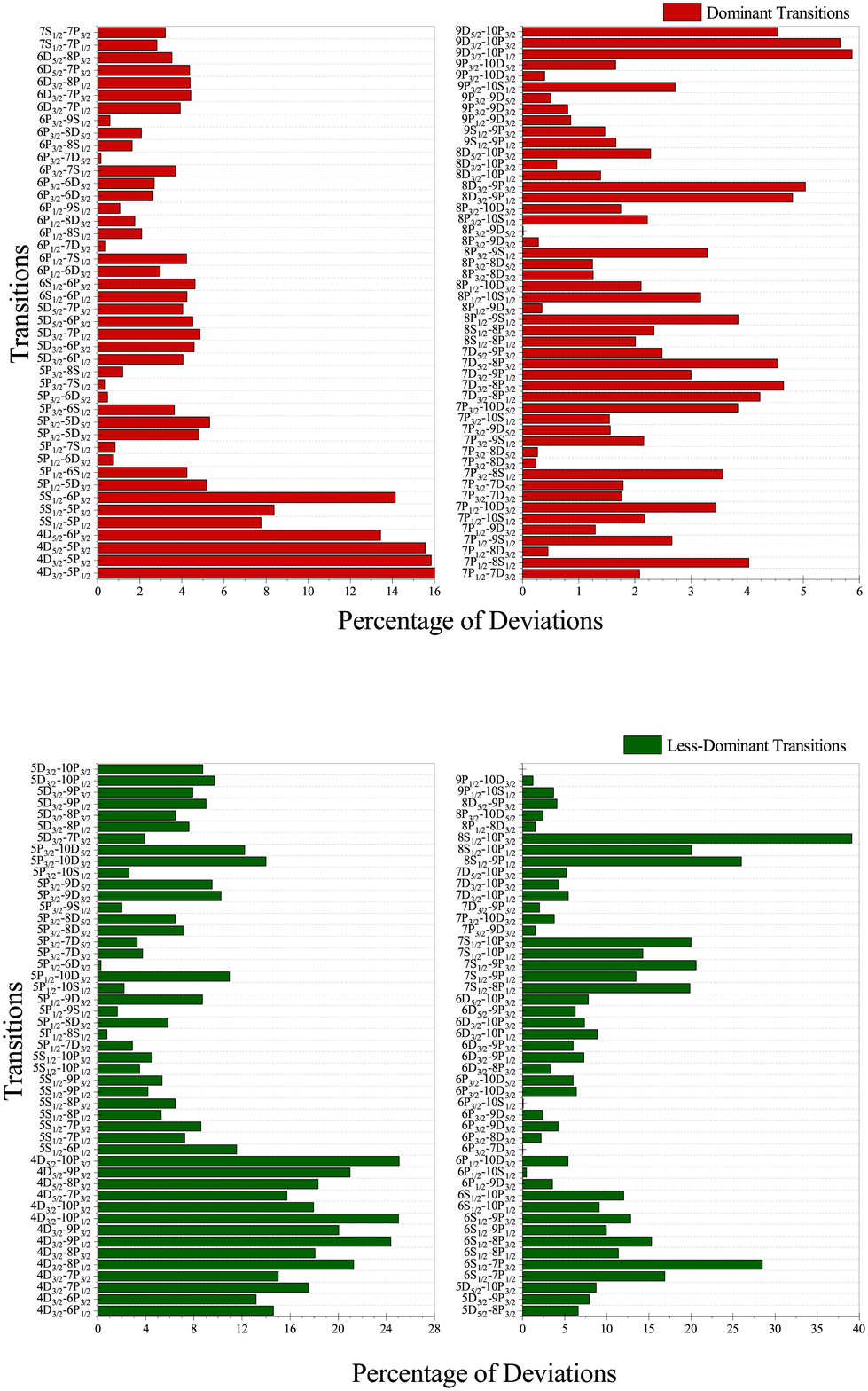}
\caption{Percentage deviations of the oscillator strengths between the length (L) and velocity (V) gauge values of the E1 transitions 
in Zr IV. Dominant and less-dominant transitions are shown separately.}
\label{zrfig1}
\end{figure}
\unskip

\clearpage
\newpage
\begin{figure}
\centering
\includegraphics[width=\textwidth,height=20cm]{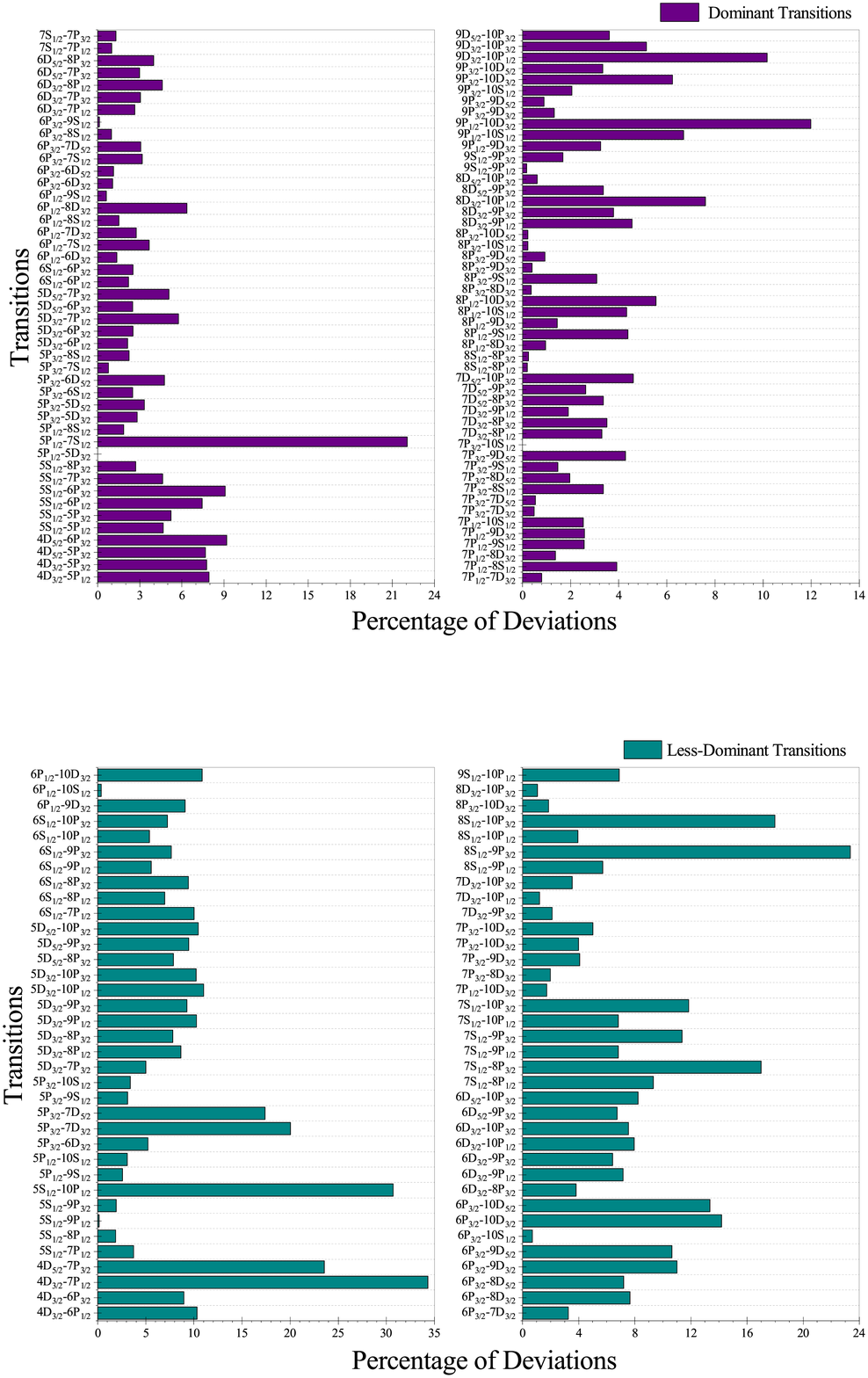}
\caption{Percentage deviation of oscillator strengths between the L and V-gauge calculations in Nb V with dominantly and less dominantly 
contributing transitions shown separately.}
\label{nbfig}
\end{figure}
\unskip
\clearpage
\twocolumn

\begin{table*}
\caption{\label{tab7} Comparison of the oscillator strengths for Zr IV and Nb V ions from our calculations with available theoretical data. 
Uncertainties are given in the parentheses.}
\begin{tabular}{@{\extracolsep\fill}ccccccc@{}}
\hline 
{Transition} & {Ion} & \multicolumn{5}{c}{$f_{kv}$} \\
 & & Present & \citep{migdalek1979relativistic} & \citep{zilitis2007oscillator} & \citep{migdalek2016relativistic} & \citep{das2017electron}\\ 
\hline
$4D_{3/2}\rightarrow 5P_{1/2}$ & Zr IV & $0.127(20)$ & & $0.157$ & $0.264$ & $0.132$\\
& Nb V & $0.132(10)$ & & $0.161$ & $0.266$ & $0.137$\\
$4D_{3/2}\rightarrow 5P_{3/2}$ & Zr IV & $0.0252(38)$ & & $0.0308$ & $0.0260$ & $0.026$\\
& Nb V & $0.0258(20)$ & & $0.0311$ & $0.0258$ & $0.027$\\
$4D_{5/2}\rightarrow 5P_{3/2}$ & Zr IV &$ 0.154(23)$ & & & $0.236$ & $0.160$\\
& Nb V & $0.158(12)$ & & & $0.235$ & $0.164$\\
$5S_{1/2}\rightarrow 5P_{1/2}$ & Zr IV & $0.312(24)$ & $0.307$ & $0.393$ & $0.331$ & $0.327$\\ 
& Nb V & $0.300(14)$ & $0.297$ & $0.385$ & $0.326$ & $0.314$\\
$5S_{1/2}\rightarrow 5P_{3/2}$ & Zr IV & $0.660(54)$ & $0.652$ & $0.829$ & $0.699$& $0.692$\\ 
& Nb V & $0.642(33)$ & $0.638$ &  $0.821$ & $0.697$ & $0.673$\\
\hline
\end{tabular}
\end{table*}

\begin{table*}
\caption{\label{tab9} The estimated lifetimes $\tau$ (in ns) for a few excited states in Zr IV and Nb V ions and their comparisons with the available 
literature data. Uncertainties are given in the parentheses.}
\begin{center}
\begin{tabular}{p{1.25in}lcc}
\hline
State & Zr IV & Nb V\\ 
\hline
$5P_{1/2}$ & $0.65(9)$ & $0.29(2)$ \\
  & $0.588$~\citep{zilitis2007oscillator} & $0.249$~\citep{zilitis2007oscillator}\\ 
& $0.622$~\citep{das2017electron} & $0.274$~\citep{das2017electron}\\
$5P_{3/2}$ & $0.61(8)$  & $0.27(2)$ \\
& $0.550$~\citep{zilitis2007oscillator} & $0.238$~\citep{zilitis2007oscillator}\\
& $0.5786$~\citep{das2017electron} & $0.259$~\citep{das2017electron}\\
$5D_{3/2}$ & $0.54(1)$ & $0.33(9)$ \\
$5D_{5/2}$ & $0.57(3)$ & $0.35(1)$ \\
$6S_{1/2}$ & $0.51(1)$ & $0.26(1)$ \\
$6P_{1/2}$ & $1.40(13)$ & $0.54(4)$ \\
$6P_{3/2}$ & $1.50(16)$ & $0.53(3)$ \\
$7S_{1/2}$ & $0.72(1)$ & $0.35(1)$ \\
$7P_{1/2}$ & $2.46(23)$ & $0.92(10)$ \\
$7P_{3/2}$ & $2.47(23)$ & $0.92(9)$ \\
$8S_{1/2}$ & $0.66(1)$ & $0.51(1)$ \\
\hline
\end{tabular}
\end{center}
\end{table*}

\section{Results and Discussion}
\label{4}

The detailed analysis of our results is given in this section along with the comparison of theoretical and experimental values available in the 
literature. First, we discuss the energies, line strengths, transition probabilities and oscillator strengths of all the transitions in both the ions. 
{The uncertainties (quoted in the parentheses) for the $S_{vk}$, $A_{vk}$  and $f_{kv}$ values have been evaluated using the uncertainties in the E1 
matrix elements, which are taken as the differences between the values from both the gauge expressions. We also depict percentage deviations of the 
oscillator strengths obtained using the V-gauge expression with respect to the L-gauge calculations for the important transitions of both the 
considered ions for better understanding of their differences.} Then, the lifetimes of the low-lying states of these ions are estimated by using the 
above transition probabilities.


{Our RMBPT calculated energy values for a few low-lying and excited states are given in Table \ref{tab10} along with their comparison with the 
experimental values listed in NIST AD ~\citep{ralchenko2008nist}. It is observed that the deviation of energy is maximum for the $4D_{3/2}$ state, 
viz, $\sim11\%$ and $7\%$, respectively, in the Zr IV and Nb V ions. However, as we move on to less penetrating states, the deviation decreases 
rapidly up to $\sim5\%$ and $3\%$ for the $8S_{1/2}$ state in Zr IV and Nb V respectively, despite the slow convergence of perturbation expansion, 
thereby increasing the accuracy of our data. Besides, the energy values for the $7D$ states of Zr IV are not provided in the NIST AD. Therefore, 
the deviations of the energy values of these states could not be estimated.}


We have listed our results for wavelengths $\lambda$, line strengths $S_{vk}$, transition probabilities $A_{vk}$ and absorption oscillator strengths 
$f_{kv}$ for Zr IV in Table \ref{tab3}. The $\lambda$ values are estimated by using the experimental energies from the NIST AD 
~\citep{ralchenko2008nist} wherever available. Otherwise, they are evaluated using our RMBPT method. Values from both the $L$ and $V$ gauge 
expressions are presented for the line strengths $S_{vk}$ in the same table. 

For the $9P_{3/2}$--$8S$ transition, we found that the reduced E1 matrix element is very small due to cancellation of large RPA contributions with
the DF value. Correspondingly, the uncertainty in this particular transition is found to be very large to consider it reliable. Hence, we have not 
considered it for further spectroscopic analysis. During the investigation of data, it has been observed that the uncertainties in the $A_{vk}$ and 
$f_{kv}$ values are mostly small for many of the considered transitions except for the $4D_{3/2}$--$(6,8,9)P_{1/2}$, $4D_{3/2,5/2}$--$8P_{3/2}$, 
$5P_{3/2}$--$10D_{3/2}$, $6S_{1/2}$--$8P_{1/2}$, $7S_{1/2}$--$8P_{1/2,3/2}$, $(7,8)S_{1/2}$--$9P_{3/2}$, $(7,9)S_{1/2}$--$10P_{3/2}$ and 
$9S_{1/2}$--$10P_{1/2}$ transitions. The large errors in these transitions are the consequence of unusually large electron correlation effects 
exhibited by the high-lying states.

{We have presented the percentage deviations of $f_{kv}$ obtained using the two gauge expressions for different transitions in Fig. \ref{zrfig1}. 
It is seen from these plots that deviation varies from $0$ to $8.5\%$ for maximum number of dominantly contributing transitions, however a maximum 
deviation of $16\%$ is seen for the $4D_{3/2}$--$5P_{1/2}$ transition. Oscillator strengths for less dominantly contributing transitions deviate majorly
between $0$--$17\%$ and in the range $21$--$29\%$ for $4D_{3/2}$--$(9,10)P_{3/2}$ and $6S_{1/2}$--$7P_{3/2}$ transitions. However, a maxima of 
$\sim39\%$ is observed for the $8S_{1/2}$--$10P_{3/2}$ transition. It is also analyzed that the $7S$--$8P_{3/2}$ transition shows an unreasonable 
percentage deviations of oscillator strengths for both the gauge expressions.}

We have tabulated our results for Nb V in Table \ref{tab4}. The $\lambda$ values are obtained by using energy values available in the NIST database
for many of the transitions; otherwise they are estimated using our RMBPT calculations. Likewise Zr IV, the RPA and BO correlation contributions are 
found to be the dominant corrections towards the final values of dipole transition amplitudes. On account of our values of the calculated E1 matrix 
elements, we have evaluated $S_{vk}$ values using both the $L$ and $V$ expressions and their uncertainties are estimated from the differences in 
in these values. The $A_{vk}$ and $f_{kv}$ values along with the uncertainties are obtained using these line strengths. It has been observed from our 
calculations that the uncertainties in the values of transition probabilities and oscillator strengths are considerably small for maximum transitions 
except for the $4D_{3/2}$--$(7,9,10)P_{3/2}$, $4D_{3/2}$--$(8-10)P_{1/2}$, $4D_{5/2}$--$(9,10)P_{3/2}$, $5S_{1/2}$--$10P_{3/2}$, 
$5P_{1/2,3/2}$--$(8-10)D_{3/2}$, $5P_{3/2}$--$(8-10)D_{5/2}$, $6S_{1/2}$--$7P_{3/2}$, $6P_{3/2}$--$10D_{5/2}$, $8P_{3/2}$--$8D_{5/2}$ and 
$9S_{1/2}$--$10P_{3/2}$ transitions. Among them, the results for the $4D_{3/2}$--$10P_{1/2}$, $5P_{1/2}$--$9D_{3/2}$ and $5P_{3/2}$--$9D_{5/2}$ 
transitions are found to be highly unreliable. Therefore, we have not provided further spectroscopic data for these three transitions specifically. 
We find that due to strong core-polarization effects arising through RPA causes such large uncertainties in the $4D_{3/2}$--$(7,9,10)P_{3/2}$, 
$4D_{3/2,5/2}$--$(8-10)P_{1/2}$, $4D_{5/2}$--$(9,10)P_{3/2}$ and $5S_{1/2}$--$10P_{3/2}$ transitions.

{We have presented plots for percent deviation of oscillator strengths of Nb V between the values obtained using both the gauge expressions and are shown
in Fig. \ref{nbfig}. In this figure, we separately present the values for dominantly and less dominantly contributing transitions. We observed that all 
the dominant transitions possess deviation less than $12\%$ except for $5P_{1/2}$--$7S_{1/2}$ transition whose deviation lies at $22\%$. For the 
other transitions, deviations of less than $20\%$ are observed for most of the transitions whereas a countable transitions have deviated from 
the L-gauge values in the range $23$--$35\%$. It is also perceived that the $4D_{3/2,5/2}$--$8P_{3/2}$, $5P_{1/2}$--$6S_{1/2}$ and 
$5P_{1/2}$--$(6,7)D_{3/2}$ transitions show unusually large percentage deviations in the oscillator strengths.}

We have also compared our results of the oscillator strengths and lifetimes of a few excited states of Zr IV with those are already available theoretical data 
in the literature for both the ions. Comparison of the oscillator strengths for the $4D_{3/2}\rightarrow 5P_{1/2,3/2}$, $4D_{5/2}\rightarrow 5P_{3/2}$
and $5S_{1/2}\rightarrow 5P_{1/2,3/2}$ transitions are made in Table \ref{tab7}. In this table, it is seen that the results for the transition 
probabilities and absorption oscillator strengths agree well with each other within the quoted error limits. We have found that our results are in 
perfect accord with the previous data published by Migdalek and Baylis ~\citep{migdalek1979relativistic} and are in fair agreement with the results
reported by Migdalek in~\citep{migdalek2016relativistic} and Das ~et.al~\citep{das2017electron} except for the $4D_{3/2}\rightarrow 5P_{1/2}$ and 
$4D_{5/2}\rightarrow 5P_{3/2}$ transitions. However, our results do not support the values obtained by Zilitis in \citep{zilitis2007oscillator}. 
This is so because, the results published in \citep{zilitis2007oscillator} are obtained by employing the mean-field calculation at the DF level. 
We have seen in our calculations that the core-polarization as well as from other effects are contributing strongly to these transition properties, 
which were neglected by Zilitis. This is why disparities between the results from both the works are seen. 

Similarly, we have compared the absorption oscillator strengths of various transitions of Nb V in Table \ref{tab7}, according to which, our results 
are in good agreement with the results given by Das ~et.al. ~\citep{das2017electron}. A deviation of less than $10\%$ is seen during the comparison of
our results with the data published in \citep{migdalek2016relativistic}, except for the $4D_{3/2}\rightarrow 5P_{1/2}$ and 
$4D_{5/2}\rightarrow 5P_{3/2}$ transitions. Our results show a variation of about $10-22\%$ with respect to the theoretical data reported by 
Zilitis ~\citep{zilitis2007oscillator} from the DF calculations. 

The estimated lifetime values of the $5P_{1/2,3/2}$, $5D_{3/2,5/2}$, $6S_{1/2}$, $6P_{1/2,3/2}, 7S_{1/2}, 7P_{1/2,3/2}$ and $8S_{1/2}$ states of Zr IV
have been listed in Table \ref{tab9}. Comparison of our calculated values for the $5P_{1/2,3/2}$ is also made in the same table. We notice that our 
results are in better agreement with the relativistic calculations presented by Das ~et.al.~\citep{das2017electron} as compared to the values given by
Zilitis \citep{zilitis2007oscillator}. The lifetimes for the $5P_{1/2,3/2}, 5D_{3/2,5/2}, 6S_{1/2}, 6P_{1/2,3/2}, 7S_{1/2}, 7P_{1/2,3/2}$ and 
$8S_{1/2}$ states of Nb V are also given in Table \ref{tab9}. These values for the $5P_{1/2,3/2}$ states show reasonable agreement with available 
other theoretical data. 

We believe that our aforementioned estimated values for various spectroscopic data are more reliable. Since the previously reported data do not 
quote the uncertainties in their calculations, our reported values will be useful in analysing various astrophysical processes involving the 
Zr IV and Nb V ions. Moreover, our precisely calculated values will be able to guide the future experiments and astrophysical observations 
to detect these spectroscopic properties.

\section{Conclusion}
\label{5}

 By evaluating electric dipole matrix elements precisely, we have determined oscillator strengths, transition probabilities and lifetimes of many 
atomic states of the rubidium-isoelectronic zirconium and niobium ions. Calculations of the matrix elements were performed by accounting  
electron correlation effects through random-phase approximation, Br\"uckner orbitals, structural radiation and normalization of wave functions
over the mean-field values from the Dirac-Fock method. Our transition properties data include $192$ transitions for Zr IV and $190$ transitions for
Nb V ion, while lifetimes are reported only for a few low-lying excited states in both the ions. We have also compared our results with the 
previously reported values for a few selected transition and find a reasonably good agreement among them except in some cases. These data with 
the quoted uncertainties can be useful for many astrophysical applications and for their their observations in the future.

\section*{Acknowledgement}

The work of B.A. is supported by SERB-TARE(TAR/2020/000189), New Delhi, India. The employed relativistic many-body method was developed in the group 
of Professor M. S. Safronova of the University of Delaware, USA.

\section*{Data availability}

The data underlying this article are available in the article.

\bibliographystyle{mnras}

\bibliography{refnb24.bib}

\bsp	
\label{lastpage}
\end{document}